\documentclass[journal]{IEEEtran}
\usepackage{cite}
\ifCLASSINFOpdf
   \usepackage[pdftex]{graphicx}
\else
   \usepackage[dvips]{graphicx}
\fi

\usepackage{amsmath}
\usepackage{amsfonts}
\usepackage{amssymb}

\usepackage{algorithmic}

\usepackage{array}

\ifCLASSOPTIONcompsoc
  \usepackage[caption=false,font=normalsize,labelfont=sf,textfont=sf]{subfig}
\else
  \usepackage[caption=false,font=footnotesize]{subfig}
\fi

\usepackage{url}

\usepackage[table,dvipsnames]{xcolor}
\usepackage[normalem]{ulem}
\useunder{\uline}{\ul}{}
\usepackage{makecell}
\usepackage{multirow}
\usepackage{enumitem}
\usepackage[colorlinks=true,breaklinks]{hyperref}
\usepackage{orcidlink}
\usepackage{booktabs}
\usepackage{tabularx}
\usepackage{mathtools}
\usepackage{placeins}
\usepackage{threeparttable}
\usepackage{microtype}
\usepackage{supertabular}

\allowdisplaybreaks
\DeclareMathOperator*{\argmax}{arg\,max}

\DeclareMathOperator{\Proj}{Proj}

\begin{document}
\bstctlcite{IEEEexample:BSTcontrol}
%
\newcommand{\mytitlename}{COMET: Concept Space Dissection of the Modality Gap in Audio-Text Multimodal Contrastive Embeddings}
\title{\mytitlename}
%
%

\author{Yonggang~Zhu \orcidlink{0009-0003-5933-8049},
        Liting~Gao \orcidlink{0009-0004-6659-382X} \IEEEmembership{Student Member, IEEE},
        Aidong~Men \orcidlink{0000-0001-6168-9276},
        Wenwu~Wang \orcidlink{0000-0002-8393-5703}
        \IEEEmembership{Fellow, IEEE}
\thanks{Y. Zhu, and A. Men are with School of Artificial Intelligence, Beijing University of Posts and Telecommunications, Beijing 100876, China. E-mail: \{zhuyonggang, menad\}@bupt.edu.cn. 

L. Gao and W. Wang are with the Centre for Vision, Speech, and Signal Processing (CVSSP), University of Surrey, GU2 7XH Guildford, U.K. E-mail: \{l.gao, w.wang\}@surrey.ac.uk.}
\thanks{This work was supported by China Scholarship Council (202506470003) and conducted during Y. Zhu's academic visit at CVSSP.}
\thanks{(Corresponding Author: Wenwu~Wang)}
}

\maketitle


\begin{abstract}

\looseness=-1 Contrastive Language-Audio Pretraining (CLAP) models are widely used for audio understanding and support modality-agnostic condition swapping in many zero-shot applications. However, their performance is heavily affected by the modality gap between audio and text embeddings. Existing explanations mainly attribute this gap to the cone effect, treating it as a shift between mean embeddings, yet correcting the mean alone yields only limited improvements. Alternative hypotheses, such as information imbalance and dimensionality collapse, have also been proposed, but they remain insufficiently verified and have not been thoroughly studied in the audio domain.
Meanwhile, several works attempt to decompose multimodal contrastive embeddings into interpretable concepts, but none explicitly analyze the modality gap from the perspective of concept decomposition. In this work, we introduce COMET (Concept space Organization and Modality gap Explanation with PLS-SVD Transformation), a novel partial least squares singular value decomposition (PLS-SVD) framework for CLAP that unveils a broader perspective of the modality gap. 
Our framework reveals that only a small, interpretable subset of axes, which captures shared concepts, contributes substantially to similarity computation, and that the mean component represents only partially the modality gap.
Building on this insight, we propose a simple spectral truncation method that mitigates the modality gap in a training-free manner. The method enables zero-shot audio captioning with condition swapping to approach fully supervised performance, without requiring large auxiliary memory banks or expensive computation. At the same time, it achieves substantial embedding dimensionality reduction while preserving strong performance on retrieval and audio captioning tasks.

\end{abstract}


%
\IEEEpeerreviewmaketitle

\section{Introduction}
\IEEEPARstart{M}{ultimodal} \looseness=-1 contrastive learning (MMCL) has achieved considerable success in recent years. These models can produce rich, transferable semantic representations that align complex data from different modalities into a shared embedding space \cite{radford2021learning,elizalde2024natural}. 
Among them, a prominent area that has attracted increasing attention is contrastive language-audio pretraining (CLAP) \cite{elizalde2024natural,wu2023large,mei2024wavcaps,sun2024auto,bai2025audiosetcaps}, where text and audio are aligned in a joint latent space. 
The CLAP models play a significant role in many audio understanding and generation tasks, including audio-text retrieval \cite{Primus2024_t8,Kim2025_t6}, audio captioning \cite{ghosh2024recap,zhu2025diffusion,zhu2026zero}, and text-guided audio generation \cite{liu2023audioldm,liu2024audioldm2}. They significantly reduce training costs and annotation efforts for these tasks through encoder reusing \cite{mei2024wavcaps,zhu2025diffusion,ghosh2024recap} and modality-agnostic condition swapping\cite{Kouzelis2023,liu2023audioldm}. The encoder reusing refers to the paradigm that after the encoders are pretrained on large-scale, noisy web data, they can be reused as a key component in many different audio understanding and generation models and attain strong performance without re-training. 

\looseness=-1 The modality-agnostic condition swapping, on the other hand, is a unique feature enabled by MMCL models like CLAP, which is of particular interest to us. This is based on the assumption that the audio and text encoders in CLAP map audio and text inputs into a shared embedding space. Consequently, when a model uses one of these embeddings as a conditioning input, it cannot distinguish whether the condition originates from the audio encoder or the text encoder. To illustrate, we briefly describe text-conditioned training for zero-shot audio captioning \cite{Kouzelis2023,deshmukh2024training,zhang2024zero,li2025drcap,zhu2026zero}, where the goal is to build a system that generates a caption, i.e. text description, for an audio clip.
In fully supervised training \cite{ghosh2024recap,zhu2025diffusion}, audio serves as the conditioning input while the caption acts as the supervisory signal. For text-conditioned training, the caption can instead be passed through the CLAP text encoder to obtain a text embedding, which is then used to condition the generation model, while the same caption remains the supervision target. Under this setup, training relies solely on text data. During inference, the conditioning input is replaced with audio embeddings, and the model is expected to treat them equivalently. This approach has been widely adopted in audio understanding and generation tasks, including audio captioning \cite{Kouzelis2023,deshmukh2024training,zhang2024zero,li2025drcap,zhu2026zero} and text-based audio generation \cite{liu2023audioldm}. By leveraging shared embedding spaces, the method enables efficient learning of cross-modal tasks from unimodal data, improving both data efficiency and adaptability across domains.

Recent research has observed that, although the text and audio embeddings produced by CLAP models are expected to lie in a shared embedding space due to contrastive pretraining, substantial discrepancies remain between the two modalities. Such discrepancies are commonly referred to as the ``modality gap'' in contrastive embeddings~\cite{liang2022mind}, which severely degrades the performance of systems relying on condition swapping \cite{Kouzelis2023,deshmukh2024training,zhang2024zero,li2025drcap,zhu2026zero}. The modality gap is often attributed to the ``cone effect''~\cite{liang2022mind}, whereby text and audio embeddings occupy two narrow yet distinct cones within the embedding space. In prior work, the offset between the cone centroids (i.e., the mean embeddings) has typically been regarded as the primary and controllable source of the gap \cite{liang2022mind}. However, our findings suggest that centroid mismatch captures only part of the overall phenomenon. Alternative explanations, including information imbalance \cite{schrodi2025two} and dimensionality collapse \cite{yi2025decipher}, have also emerged to understand the modality gap in contrastive language image pretraining (CLIP)~\cite{radford2021learning}. Nevertheless, these explanations remain fragmented and sometimes lack comprehensive experimental validation. 

\looseness=-1 Existing approaches for bridging the gap in audio understanding and generation can broadly be divided into training-free and training-based methods. Representative training-free approaches include embedding shift (ES) \cite{Kouzelis2023,liang2022mind}, Gaussian noise injection (NI) \cite{Kouzelis2023,deshmukh2024training,zhang2024zero,zhu2026zero}, and projection decoding (PD) \cite{li2023decap,Kouzelis2023,li2025drcap}. These methods operate directly on off-the-shelf CLAP models and map source-modality embeddings into the target-modality space without training additional mapping networks, making them computationally efficient and broadly applicable. Training-based approaches, on the contrary, typically introduce additional learnable modules~\cite{mo2025diffgap,nam2025diffusion} or re-train a CLAP model~\cite{sofer2025pull}. For example, the methods in \cite{mo2025diffgap,nam2025diffusion} employ diffusion models trained on paired audio-text embeddings to learn the cross-modal mapping in a fully supervised manner, while the method~\cite{sofer2025pull} trains new MMCL models with auxiliary objectives to explicitly reduce the modality gap.
Compared with training-based methods, training-free techniques are simpler and substantially more computationally efficient, and are therefore the primary focus of this work. We propose a new training-free method with improved memory usage and competitive performance, which is introduced later in this part.

\looseness=-1 Moreover, several studies attempt to interpret MMCL embeddings by decomposing them into combinations of human-interpretable concepts \cite{NEURIPS2024_splice,zhao2025quantifying,gordon2026scocca}. However, these approaches do not explicitly examine the structure of the embedding space from a multimodal perspective, nor do they investigate how the modality gap can be bridged within their frameworks. In contrast, we propose a new, efficient concept decomposition framework for CLAP that addresses these issues in a unified manner.

\looseness=-1 In this work, we present COMET (Concept space Organization and Modality gap Explanation with PLS-SVD Transformation), which is a unified framework for understanding CLAP. It reveals the connections among concept decomposition, multimodal similarity computation, and the various sources of the modality gap. Specifically, we introduce a concept-aware, bottom-up decomposition of the CLAP latent space based on Partial Least Squares Singular Value Decomposition (PLS-SVD), which naturally captures both multimodal alignment and concept saliency. 
The resulting decomposition reveals a comet-like latent structure: a compact, luminous \textbf{head} of shared cross-modal semantics, and a long, diffuse, modality-private \textbf{tail}. 
This framework enables a detailed examination of the intrinsic ``native language'' of CLAP without relying on external bases, nonlinear transformations, or black-box models. Using this perspective, we derive a theoretically and empirically grounded interpretation of projection decoding, which is a well-performing method for reducing modality gap \cite{li2023decap,Kouzelis2023,li2025drcap} that has not been well understood, providing new insights into the underlying causes of the modality gap. Building on these findings, we further propose a fast, training-free correction method for CLAP embeddings that offers several advantages. First, it effectively mitigates the modality gap, enabling competitive zero-shot audio captioning performance comparable to fully supervised approaches, while eliminating the need for the large memory banks required by projection decoding \cite{Kouzelis2023}. Second, the proposed method substantially reduces embedding dimension while maintaining, and in some cases improving, performance on retrieval and audio captioning tasks. Our main contributions are:

\begin{itemize}
    \item \looseness=-1 We propose a PLS-SVD framework to analyse the CLAP embeddings, and identify a rank-structured nature of the CLAP latent space that consists of a \textbf{mean component}, a compact \textbf{shared semantic head}, and a long \textbf{modality-private tail}. While the ``cone effect'' focuses on the mean component, we find that each part acts as different sources of the modality gap. This allows the reduction of the gap in a principled way. 
    \item We find that projection decoding is a \textbf{concept space truncation and imputation} filter on top of mean shifting.
    \item \looseness=-1 We propose PLSHead, a fast, versatile, memory-efficient spectral truncation method that strongly compresses the embeddings, closes the modality gap, and achieves comparable or even better performance on downstream tasks such as audio-text retrieval and audio captioning.
    
\end{itemize}

\section{Related Work}
\subsection{Audio-Text Contrastive Learning}
\looseness=-1 Audio-text contrastive learning aims to map audio and text into a joint embedding space, utilizing loss functions like InfoNCE \cite{he2020momentum} to pull matched representations together while pushing unmatched pairs apart.
Inspired by CLIP \cite{radford2021learning}, the scale of paired audio-text data for contrastive training expands over time \cite{elizalde2024natural,wu2023large,mei2024wavcaps,sun2024auto,bai2025audiosetcaps}, leading to the development of CLAP models. They usually use a pipeline to automatically expand data. Typical techniques include web scraping \cite{wu2023large}, generating captions from metadata with large language models (LLM)~\cite{mei2024wavcaps}, or combining detection models and LLMs to extract contents and expand them into captions altogether~\cite{sun2024auto,bai2025audiosetcaps}. 
They show strong performance across a range of downstream tasks, such as audio-text retrieval~\cite{mei2024wavcaps,sun2024auto,bai2025audiosetcaps}, audio captioning~\cite{mei2024wavcaps,sun2024auto,bai2025audiosetcaps}, and zero-shot audio classification~\cite{mei2024wavcaps,sun2024auto,bai2025audiosetcaps}. 
They function widely as retrieval models \cite{Kim2025_t6,ghosh2024recap,li2025drcap}, general-purpose audio encoders \cite{deshmukh2023pengi,liu2023audioldm}, and evaluators of generation quality \cite{gao2025rfm}. In addition, they enable modality-agnostic condition swapping, which facilitates many zero-shot applications~\cite{Kouzelis2023,deshmukh2024training,zhang2024zero,li2025drcap,zhu2026zero,liu2023audioldm}.

\subsection{Post-Hoc Concept Decomposition for MMCL}

\looseness=-1 Post-hoc concept decomposition methods are widely used for interpreting neural networks by expressing internal representations as combinations of human-understandable concepts \cite{fel2023holistic}. Representative approaches such as TCAV \cite{kim2018interpretability}, CRAFT \cite{fel2023craft}, and MCD \cite{vielhaben2023multi} focus primarily on single-modality activations. Another line of work builds on concept bottleneck models (CBMs) \cite{koh2020concept,oikarinen2023label,yuksekgonul2023posthoccbm}, which learn intermediate concepts and map them to target classes. However, these methods typically require manually curated concepts or externally constructed vocabularies, and mainly target supervised image classification rather than multimodal contrastive spaces.
For multimodal embedding spaces, SpLiCE \cite{NEURIPS2024_splice} sparsely projects image embeddings onto a massive dictionary of text-derived concepts, an approach later extended to CLAP \cite{zhang2025transformation}. Similarly, the method \cite{zang2025discoverablevisualconcept} constructs concept vocabularies using LLM-generated phrases. While intuitive, these top-down approaches rely on predefined external vocabularies and may discard substantial embedding information \cite{zhao2025quantifying}.

\looseness=-1 More recently, the method in \cite{zhao2025quantifying} applies singular value decomposition (SVD) and Varimax rotation to identify unimodal concepts, but does not explicitly model cross-modal interactions or the alignment objective. The method in \cite{gordon2026scocca} combines canonical-correlation analysis (CCA), Hungarian matching, and iterative optimization to interpret CLIP embeddings, but this method is computationally expensive and depends on whitening operations that remove raw signal strength, an important indicator of semantic saliency. 
In contrast, our method provides a unified and efficient analysis of the CLAP embedding space. Rather than relying on predefined vocabularies or complex optimization pipelines, we adopt a bottom-up decomposition that automatically discovers semantic axes directly from audio-text alignment. Our framework jointly captures alignment and signal strength, identifies the components contributing most to multimodal similarity, visualizes the modality gap, explains projection decoding, and enables effective gap mitigation while preserving downstream performance with only a compact set of bases.

\subsection{Modality Gap in MMCL}\label{sec:rel_work_mod_gap}
\looseness=-1 
The modality gap in MMCL embeddings was first identified by Liang et al.~\cite{liang2022mind}, who attributed it to the ``cone effect'': image and text embeddings occupy different cones after random encoder initialization, and these cones persist throughout contrastive training. The gap is usually quantified by the distance between cone centroids (mean embeddings). This perspective naturally suggests that manipulating these centroids, such as by subtracting the image mean and adding the text mean for image embeddings, can force the cones to overlap, serving as the core intuition for general embedding-shift methods. Building on this cone-effect interpretation, many subsequent studies have further analyzed the modality gap~\cite{zhang2023diagnosing,shi2023towards,zhang2024connect,li2025closing,yaras2025explaining,grassucci2025closing}. For example, Shi et al.~\cite{shi2023towards} and Yaras et al.~\cite{yaras2025explaining} study how the contrastive temperature parameter affects the persistence of the gap. Zhang et al.~\cite{zhang2024connect} further model the discrepancy between paired image and text embeddings as a constant mean-offset term plus a white Gaussian noise component representing random alignment noise. In contrast, our findings indicate that this residual component contains meaningful structure beyond simple Gaussian noise, suggesting a broader interpretation than the conventional cone-effect view.

Other works propose alternative explanations for the modality gap. Schrodi et al.~\cite{schrodi2025two} attribute it to information imbalance, arguing that one modality (e.g., images) often contains substantially richer information than the other (e.g., short captions), making perfect alignment inherently difficult. However, their analysis is performed in the raw embedding space, where semantic structure and noise remain entangled. In contrast, our approach explicitly separates shared semantic components from unaligned noise, allowing us to identify where alignment breaks down. Ramasinghe et al.~\cite{ramasinghe2024accept} regard the gap as inevitable due to intrinsic modality differences, whereas our results show that the gap is structured and can be systematically reduced. Yi et al.~\cite{yi2025decipher} explain the gap through dimension collapse, though their theory relies on idealized assumptions and still characterizes the gap primarily through centroid distances. Fahim et al.~\cite{fahim2024notamodalitygap} argue that the gap is largely induced by contrastive loss and can be mitigated using additional uniformity and alignment objectives~\cite{wang2020understanding}, but this requires retraining the model.

\looseness=-1 Existing methods for reducing the modality gap can broadly be divided into ad-hoc training approaches and post-hoc methods. Ad-hoc approaches modify the MMCL training process through specialized losses, architectures, or training strategies. Examples include adding uniformity and alignment objectives~\cite{fahim2024notamodalitygap,wang2020understanding,grassucci2025closing}, adversarial domain adaptation losses~\cite{sofer2025pull}, encoder batch normalization~\cite{an2025i0t}, shared Transformer architectures with projection layers~\cite{eslami2025mitigate}, and strategies such as temperature scheduling or feature swapping~\cite{yaras2025explaining}. While effective, these approaches require retraining large-scale MMCL models such as CLIP or CLAP, which is computationally expensive.

\looseness=-1 Post-hoc methods instead operate on off-the-shelf MMCL models without modifying the original training. Some approaches still require training auxiliary models, such as diffusion-based mappings between modalities~\cite{mo2025diffgap,nam2025diffusion,lee2025diffusion}, which remain computationally demanding. Training-free post-hoc methods are therefore more lightweight and portable. Embedding-shift and centering methods~\cite{liang2022mind,Kouzelis2023,an2025i0t} align modality centroids through mean shifting and renormalization. Gaussian noise injection~\cite{nukrai2022text} blends embeddings across modalities but can obscure semantic information. Projection decoding~\cite{li2023decap} reconstructs a target-modality embedding as a weighted combination of source-modality embeddings from a memory bank. This approach has shown strong empirical performance in zero-shot captioning systems~\cite{Kouzelis2023,li2025drcap,li2023decap}. However, it depends on a large memory bank, typically constructed from all text embeddings in the training set, and its underlying mechanism remains poorly understood. In this work, we provide a theoretical explanation for projection decoding and remove the need for large memory banks.

\subsection{Zero-shot Audio Captioning}
\looseness=-1 
Audio captioning aims to generate natural language descriptions of acoustic scenes and events from an input audio clip. Zero-shot audio captioning further requires that the main captioning model be trained without paired audio--text supervision~\cite{Kouzelis2023,deshmukh2024training,zhang2024zero,li2025drcap,zhu2026zero,nukrai2022text,li2023decap,gu2023can,salewski2023zero,shaharabany2023zero}. However, paired audio--text data may still be used in preparatory stages prior to training the captioning model itself. For instance, many systems rely on pretrained CLAP models, which require paired audio--text data during pretraining. Similarly, diffusion-based embedding mapping methods~\cite{mo2025diffgap,nam2025diffusion} are trained on paired data, while embedding-shift approaches~\cite{Kouzelis2023} might use paired embeddings to estimate the centroid difference between modalities.

Existing zero-shot audio captioning methods can be broadly categorized into decoder-guided~\cite{salewski2023zero,shaharabany2023zero}, encoder-guided~\cite{Kouzelis2023,deshmukh2024training,zhang2024zero,li2025drcap,zhu2026zero,nukrai2022text,li2023decap,gu2023can}, and hybrid approaches~\cite{zhu2026zero}. Decoder-guided methods adjust the word probabilities from the decoder using CLAP-based similarity scores, but generally underperform encoder-guided approaches~\cite{zhang2024zero,zhu2026zero}. Encoder-guided methods instead rely on modality-agnostic condition swapping: during training, the condition is encoded using the CLAP text encoder, while at inference time it is replaced by embeddings from the audio encoder. To improve cross-modal alignment, these methods often incorporate post-hoc modality-gap mitigation techniques discussed in Sec.~\ref{sec:rel_work_mod_gap}. Additional enhancements include data augmentation~\cite{zhang2024zero}, keyword guidance~\cite{zhang2024zero}, and retrieval of semantically similar captions~\cite{li2025drcap}. We evaluate our method on the two standard audio captioning benchmarks, Clotho~\cite{drossos2020clotho} and AudioCaps~\cite{kim2019audiocaps}.

\section{Framework: PLS-SVD and the Subspace Structure of CLAP Embeddings}
\subsection{Notations}
Assume a CLAP model produces $N$ text-audio embedding pairs $(\boldsymbol{t}_i,\boldsymbol{a}_i),i=1\dots N$, by encoding paired text and audio data through the text and audio encoders, respectively. Here, $\boldsymbol{t}_i,\boldsymbol{a}_i\in\mathbb{R}^C$, where $C$ is the embedding dimension, such as 1024 in Mei et al. \cite{mei2024wavcaps} or 768 in Bai et al. \cite{bai2025audiosetcaps}. 
The mean text and audio embeddings are $\bar{\boldsymbol{t}}=1/N\sum_{i=1}^N\boldsymbol{t}_i$, and $\bar{\boldsymbol{a}}=1/N\sum_{i=1}^N\boldsymbol{a}_i$, respectively. The centered embeddings are $\tilde{\boldsymbol{t}}_i=\boldsymbol{t}_i-\bar{\boldsymbol{t}}$ and $\tilde{\boldsymbol{a}}_i=\boldsymbol{a}_i-\bar{\boldsymbol{a}}$.
We denote a projection axis for the text embedding as $\boldsymbol{u}\in\mathbb{R}^C$, and the projection value of $\boldsymbol{t}_i$ on $\boldsymbol{u}$ is $\Proj_{\boldsymbol{u}}(\boldsymbol{t}_i)$. Similarly, the projection of an audio embedding $\boldsymbol{a}_i$ on an audio axis $\boldsymbol{v}$ is $\Proj_{\boldsymbol{v}}(\boldsymbol{a}_i)$. In this paper, all vectors are column vectors. Datasets like Clotho \cite{drossos2020clotho} or the test set of AudioCaps \cite{kim2019audiocaps} have 5 captions for each audio. For them, each audio is replicated 5 times so that every 5 consecutive audios are the same, and $(\boldsymbol{t}_i,\boldsymbol{a}_i)$ still denotes a positive pair.
\subsection{Mathematical Formulation}
We start by examining how to find axes of key, shared semantics. We note that if, for $i=1\dots N$, the projection values $\operatorname{Proj}_{\boldsymbol{u}}({\boldsymbol{t}}_i)$ and $\operatorname{Proj}_{\boldsymbol{v}}({\boldsymbol{a}}_i)$ frequently have large variations of the same sign, this could indicate that $\boldsymbol{u}$ and $\boldsymbol{v}$ are both important directions and encode the same meaning in the text and audio spaces, respectively. In other words, if the covariance $\sum_{i=1}^N\operatorname{Proj}_{\boldsymbol{u}}(\tilde{\boldsymbol{t}}_i)\operatorname{Proj}_{\boldsymbol{v}}(\tilde{\boldsymbol{a}}_i)$ is large, the direction pair $(\boldsymbol{u},\boldsymbol{v})$ encodes core shared semantics. Centered embeddings $\tilde{\boldsymbol{t}}_i,\tilde{\boldsymbol{a}}_i$ are used since the variation of the projection can be obtained by mean subtraction before projection. Formally, we find:
\begin{equation}\label{eq:plssvd_target}
    \argmax_{\boldsymbol{u},\boldsymbol{v}}\ (T\boldsymbol{u})^T(A\boldsymbol{v})=\argmax_{\boldsymbol{u},\boldsymbol{v}}\ \boldsymbol{u}^T(T^TA)\boldsymbol{v}.
\end{equation}
Here, we stack the centered text embeddings row-wise to obtain $T\in\mathbb{R}^{N\times C}$ and centered audio embeddings to obtain $A\in\mathbb{R}^{N\times C}$. The $i$-th row in $T$ and the $i$-th row in $A$ form a matching pair of centered text-audio embeddings. This problem can be solved efficiently with Partial Least Squares Singular Value Decomposition (PLS-SVD), which performs SVD on the text-audio covariance matrix $M=T^TA$:
\begin{equation}\label{eq:plssvd_svd}
    M=U\Sigma V^T.
\end{equation}
Here, the SVD returns a ranked list of paired directions that satisfy the objective in Eq.~(\ref{eq:plssvd_target}) from strong to weak. The $i$-th column of $U\in\mathbb{R}^{C\times C}$, which is $\boldsymbol{u}_i\in\mathbb{R}^C$, and the $i$-th column of $V\in\mathbb{R}^{C\times C}$, $\boldsymbol{v}_i$, form the $i$-th text-audio direction pair. The text directions $\boldsymbol{u}_i$ are mutually orthogonal, providing a unique, complete description of a text embedding. The orthogonality allows each concept to vary freely without interfering with each other. The audio directions $\boldsymbol{v}_i$ also have these features. 
$\Sigma\in\mathbb{R}^{C\times C}$ is a diagonal matrix of singular values. Its element $\Sigma_{ii}$ is the covariance $\boldsymbol{u}_i^T(T^TA)\boldsymbol{v}_i$, measuring the degree to which the directions $\boldsymbol{u}_i$ and $\boldsymbol{v}_i$ encode core shared semantics. The first direction pair maximizes Eq.~(\ref{eq:plssvd_target}), and the second pair is the second largest following orthogonality constraints, etc. In this way, we obtain a decomposition of the embeddings as:
\begin{equation}\label{eq:emb_dissection}
    \boldsymbol{t}_i=\bar{\boldsymbol{t}}+\sum_{j=1}^C\hat{t}_{ij}\boldsymbol{u}_j,\ \boldsymbol{a}_i=\bar{\boldsymbol{a}}+\sum_{j=1}^C\hat{a}_{ij}\boldsymbol{v}_j.
\end{equation}
\looseness=-1 Here, the top directions should encode core shared semantics, and the tail directions could contain modality-specific content or noise. The terms $\hat{t}_{ij}$ and $\hat{a}_{ij}$ are the projection coefficients, which are the $i$-th elements of $T\boldsymbol{u}_j$ and $A\boldsymbol{v}_j$, respectively. The mean term, on the other hand, encodes the static modality gap noted in previous works, which is stable across samples and could arise from the cone effect and contrastive training \cite{liang2022mind,zhang2023diagnosing,shi2023towards,zhang2024connect,li2025closing,yaras2025explaining,grassucci2025closing}. Our full decomposition, as shown later, reveals other sources of the modality gap. 
Note that the size of $M\in\mathbb{R}^{C\times C}$ is irrelevant to the total number of samples $N$, and $C$ is usually relatively stable and much smaller than the ever-increasing $N$. The SVD can be computed efficiently for this size of $M$, enabling the method to scale in large-scale analyses.

\subsection{Dissecting the CLAP Embeddings}\label{sec:dissect_clap_emb}
We inspect the decomposition with the Clotho \cite{drossos2020clotho} training set and a commonly-used CLAP model, HTSAT-BERT-ZS \cite{mei2024wavcaps}, which is pretrained on WavCaps \cite{mei2024wavcaps} in a zero-shot manner.\footnote{The dissection is repeated using CLAP pretrained on other datasets (e.g. AudioSetCaps \cite{bai2025audiosetcaps}, WavCaps+SoundVECaps\cite{yuan2025sound}), and CLAP having other model structures. Both Clotho \cite{drossos2020clotho} and AudioCaps \cite{kim2019audiocaps} are tried for the dissection. The results are in App.~\ref{sec:app_decomp_ret_other_clap}. They show a similar pattern. Intriguingly, the head sizes are also roughly 100.} 
The ``zero-shot'' means all overlapping samples from Clotho \cite{drossos2020clotho} and AudioCaps \cite{kim2019audiocaps} are excluded from pretraining, better demonstrating real generalization abilities of the CLAP models. This model has an embedding dimension of $C=1024$.

\begin{figure}
    \centering
    \includegraphics[width=1.0\linewidth]{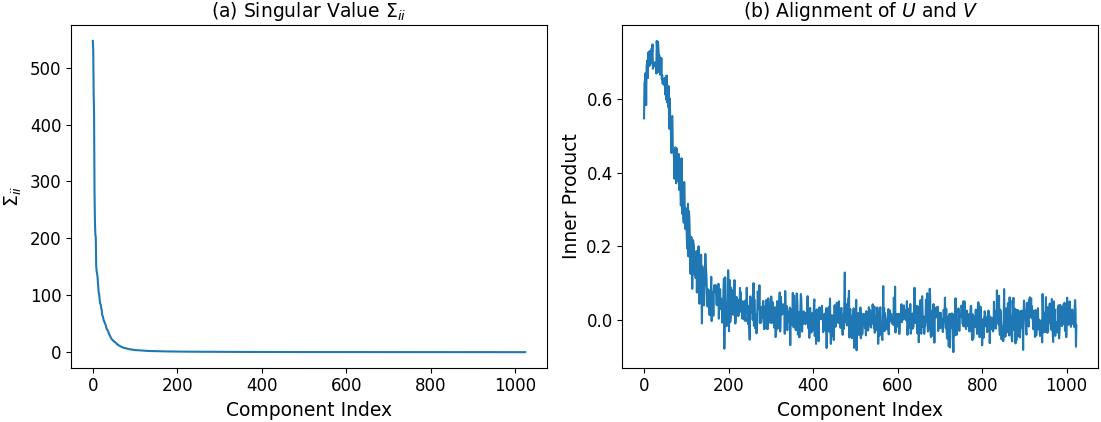}
    \caption{\textbf{The singular value $\Sigma_{ii}$ and its relationship with UV alignment $\boldsymbol{u}_i\cdot\boldsymbol{v}_i$.} (a) $\Sigma_{ii}$ drops drastically to near-zero values within the top 100 indices, revealing a small, shared core semantic head. (b) $\boldsymbol{u}_i\cdot\boldsymbol{v}_i$ briefly increases, then drops massively within the top 100 indices.}
    \label{fig:clotho_s_uv}
\end{figure}

We perform PLS-SVD on the Clotho training set according to Eq.~(\ref{eq:plssvd_svd}) and show the distribution of the singular values $\Sigma_{ii}$ in Fig.~\ref{fig:clotho_s_uv}. We also hypothesize that for the top indices, since they encode shared semantics, the text axis $\boldsymbol{u}_i$ and audio axis $\boldsymbol{v}_i$ should be well-aligned with large inner products. Also, direction pairs in the tail should have smaller inner products. We name $\boldsymbol{u}_i\cdot\boldsymbol{v}_i$ the UV alignments, which are in the second panel of Fig.~\ref{fig:clotho_s_uv}. The results show that the singular values $\Sigma_{ii}$ drop quickly within the top 100 indices and remain near-zero for the rest of the indices. The UV alignments $\boldsymbol{u}_i\cdot\boldsymbol{v}_i$ briefly increase, then sharply drop to low values within the top 100 indices. This indicates the low intrinsic dimensionality of this CLAP embedding space. It reveals \textbf{a small shared semantic head}: only a few components actively have large, shared variations. In addition, only these components have well-aligned $(\boldsymbol{u}_i,\boldsymbol{v}_i)$ pairs.

\begin{figure}
    \centering
    \includegraphics[width=0.61\linewidth]{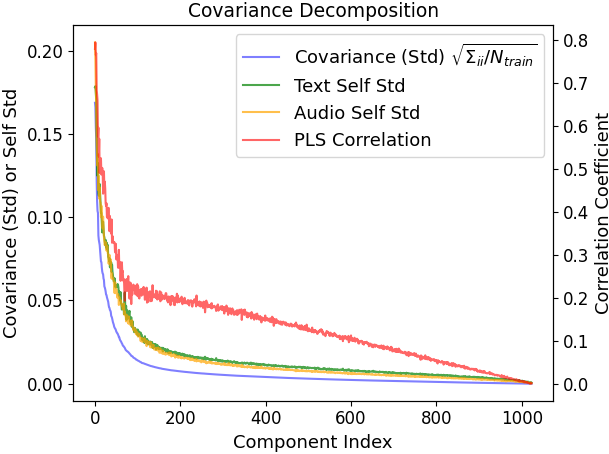}
    \caption{\textbf{Decomposition of the covariance $\Sigma_{ii}/N_{\text{train}}$.} The text and audio self-std ($\sqrt{\operatorname{Var}(T\boldsymbol{u}_i)}$ and $\sqrt{\operatorname{Var}(A\boldsymbol{v}_i)}$) decay slower than $\sqrt{\Sigma_{ii}/N_{\text{train}}}$. The correlation coefficient quickly drops to about 0.23 before index 100, enters a small transition area, and slowly, approximately linearly decays to 0.}
    \label{fig:clotho_sdecomp}
\end{figure}

How can we further explain the singular value $\Sigma_{ii}$? Recall that it is the covariance of the projections $T\boldsymbol{u}_i$ and $A\boldsymbol{v}_i$, which considers both the signal strength in each modality and their correlation. Thus, we decompose it into three parts:
\begin{equation}\label{eq:cov_dissect}
    \operatorname{Cov}(T\boldsymbol{u}_i,A\boldsymbol{v}_i)=\operatorname{Corr}(T\boldsymbol{u}_i,A\boldsymbol{v}_i)\sqrt{\operatorname{Var}(T\boldsymbol{u}_i)}\sqrt{\operatorname{Var}(A\boldsymbol{v}_i)},
\end{equation}where normalized covariance $\operatorname{Cov}(T\boldsymbol{u}_i,A\boldsymbol{v}_i)=\Sigma_{ii}/N_{\text{train}}=(T\boldsymbol{u}_i)^T(A\boldsymbol{v}_i)/N_{\text{train}}$, and normalized variance $\operatorname{Var}(\boldsymbol{x})=\boldsymbol{x}^T\boldsymbol{x}/N_{\text{train}}$ for a centered vector $\boldsymbol{x}\in\mathbb{R}^{N_{\text{train}}}$. These variances show the energy in the projection values of the text embeddings $T\boldsymbol{u}_i$ and audio embeddings $A\boldsymbol{v}_i$. We plot the square-rooted covariance $\sqrt{\Sigma_{ii}/N_{\text{train}}}$ together with $\sqrt{\operatorname{Var}(T\boldsymbol{u}_i)}$ and $\sqrt{\operatorname{Var}(A\boldsymbol{v}_i)}$, since they all denote square-rooted energy. $\operatorname{Corr}\in[0,1]$ is the correlation coefficient. The plot is in Fig.~\ref{fig:clotho_sdecomp}. 

\looseness=-1 As shown in Fig.~\ref{fig:clotho_sdecomp}, $\sqrt{\operatorname{Var}(T\boldsymbol{u}_i)}$ and $\sqrt{\operatorname{Var}(A\boldsymbol{v}_i)}$ decay much more slowly than $\sqrt{\Sigma_{ii}/N_{\text{train}}}$. This indicates that a significant portion of the self energy could reside in the less aligned tail subspaces. 
For a quantitative comparison, we calculate the mean norm of 1) the top 100 indices, 2) the remaining indices (101 to $C$), and 3) the whole centered embedding:
\begin{equation}
        \overline{\text{norm}}_t^{st:ed}=\frac{1}{N}\sum\nolimits_{i=1}^N\sqrt{\sum\nolimits_{j=st}^{ed}(\hat{t}_{ij})^2}.
\end{equation}
The definition is similar for audio embeddings. The result is $\overline{\text{norm}}_t^{1:100}=0.770,\overline{\text{norm}}_t^{101:C}=0.344,\overline{\text{norm}}_t^{1:C}=0.845$; and $\overline{\text{norm}}_a^{1:100}=0.776,\overline{\text{norm}}_a^{101:C}=0.301,\overline{\text{norm}}_a^{101:C}=0.834$. We can see that \textbf{the tail indices (101 to $C$) indeed contain a significant portion of the total energy}. 
\begin{figure}
\centering
\subfloat[The full matrix]{
    \includegraphics[width=0.47\columnwidth]{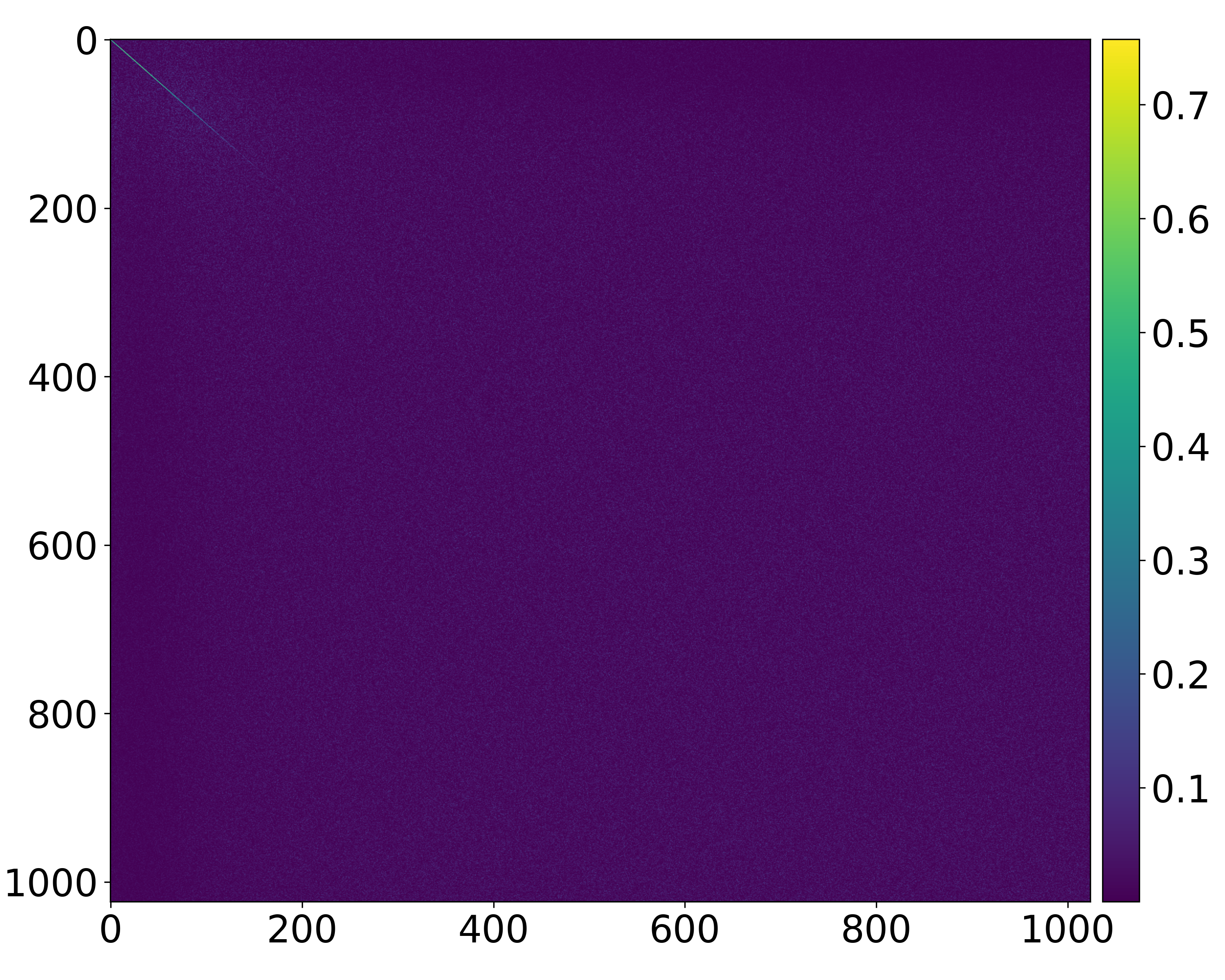}
}
\subfloat[Zoom-in of first 300$\times$300 elements]{
    \includegraphics[width=0.47\columnwidth]{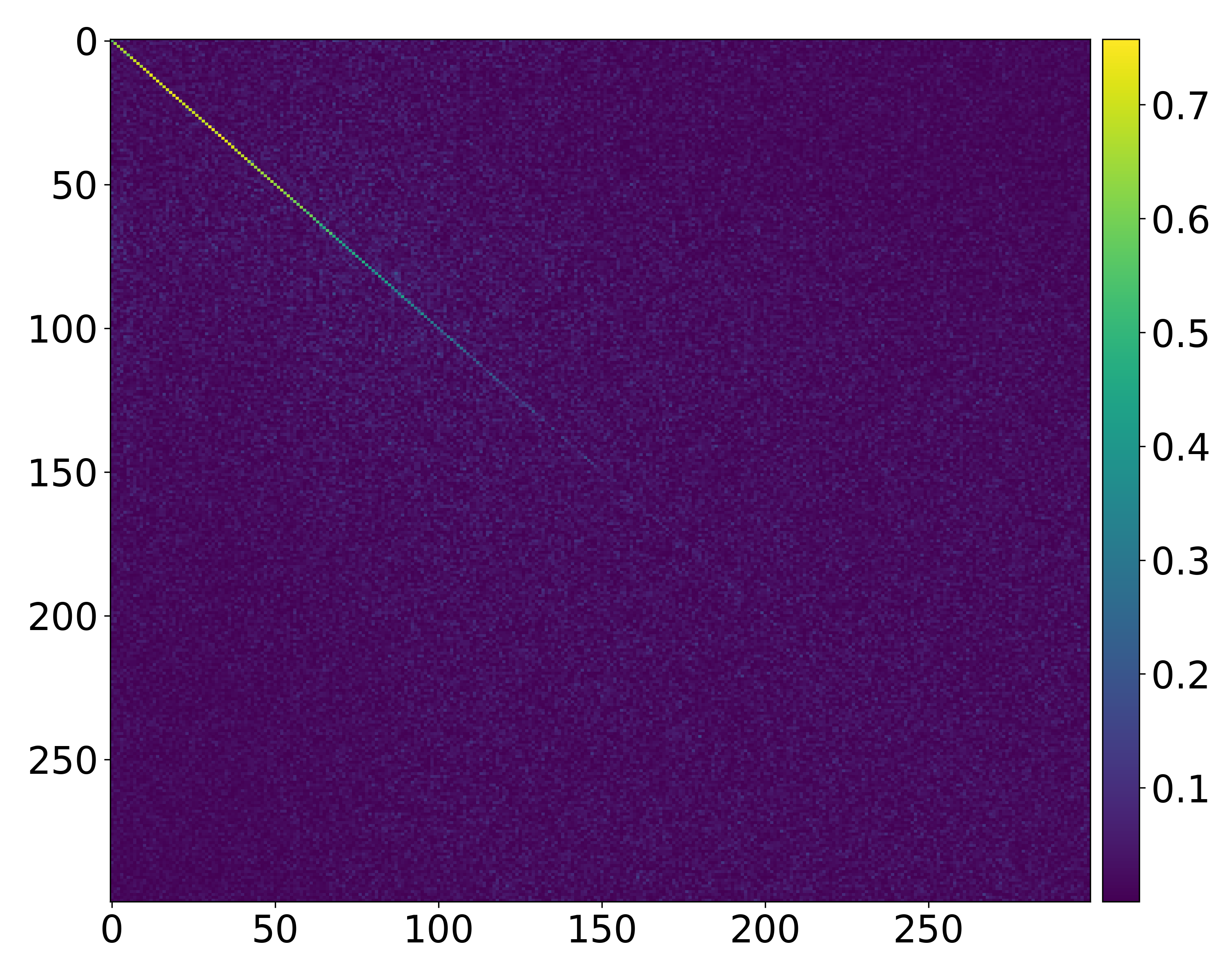}
}
    \caption{\textbf{Visualization of the absolute of the UV matrix $|U^TV|$.} The origin is at the upper-left corner. The diagonal elements in the upper-left corner have large activations, while all other parts only have very small activations.}
    \label{fig:uvmatrix}
\end{figure}

The switch from a quick drop to a slow but approximately linear decay in the correlation coefficient in  Fig.~\ref{fig:clotho_sdecomp} could indicate a transition in the nature of the head and tail subspaces. However, how can we further understand these subspaces? 
Let's develop our intuition by examining how the decomposition in Eq.~(\ref{eq:emb_dissection}) helps to explain the inner product similarity computation. Starting with a positive pair $\boldsymbol{t}_i\cdot \boldsymbol{a}_i$, we have:
\begin{align}\label{eq:simdissectfull}
\boldsymbol{t}_i\cdot \boldsymbol{a}_i&\approx (\sum\nolimits_{j=1}^{C}\hat{t}_{ij}\boldsymbol{u}_j)(\sum\nolimits_{j=1}^{C}\hat{a}_{ij}\boldsymbol{v}_j)\notag\\
&= \underbrace{\sum\nolimits_{j=1}^{C}\hat{t}_{ij}\hat{a}_{ij}(\boldsymbol{u}_j\cdot \boldsymbol{v}_j)}_{\text{direct effect}}+\underbrace{\sum\nolimits_{k\neq l}\hat{t}_{ik}\hat{a}_{il}(\boldsymbol{u}_k\cdot \boldsymbol{v}_l)}_{\text{cross effect}}.
\end{align}
\looseness=-1 We see that the similarity calculation can be divided into a ``direct effect'' component, where each text direction $\boldsymbol{u}_j$ directly acts on its matching audio direction $\boldsymbol{v}_j$, and a ``cross effect'' component, where each direction $\boldsymbol{u}_k$ interacts with a different semantic component $\boldsymbol{v}_l$. We hypothesize that the direct effects could be more important in the similarity calculation. To verify this, we first examine the UV matrix $U^TV$, where each element $(U^TV)_{ij}=\boldsymbol{u}_i\cdot\boldsymbol{v}_j$. We take the absolute value to see its magnitude. As shown in Fig.~\ref{fig:uvmatrix}, \textbf{the $|U^TV|$ is particularly bright only in the upper-left corner of the diagonal}, while all other elements only have small values. Since the UV alignments $\boldsymbol{u}_i\cdot\boldsymbol{v}_j$ are the ``weights'' in the dissected similarity calculation (Eq.~(\ref{eq:simdissectfull})), we deduce that the cross effect terms could be less important, and only the head of the direct effect terms is actively contributing to the similarity. Continuing from Eq.~(\ref{eq:simdissectfull}), we have:
\begin{equation}\label{eq:simdissectfullcont}
    \boldsymbol{t}_i\cdot \boldsymbol{a}_i
    \approx \sum_{j=1}^{C}\hat{t}_{ij}\hat{a}_{ij}(\boldsymbol{u}_j\cdot \boldsymbol{v}_j)
    \approx \sum_{j=1}^{K}\hat{t}_{ij}\hat{a}_{ij}(\boldsymbol{u}_j\cdot \boldsymbol{v}_j),
\end{equation}
where $K$ denotes the dimensionality of the shared head. 
We further verify whether the cross effect terms really contribute less significantly by computing the mean absolute contribution of the direct effect vs. the cross effect. We also compute this metric for the direct effect of the first 100 components. The formulae for positive pairs are as follows.
\begin{gather}
    \overline{\text{Contrib}}_{\text{direct effect}}=\frac{1}{N}\sum\nolimits_{i=1}^N\left\lvert\sum\nolimits_{j=1}^{C}\hat{t}_{ij}\hat{a}_{ij}(\boldsymbol{u}_j\cdot \boldsymbol{v}_j)\right\lvert,\notag\\
    \overline{\text{Contrib}}_{\text{cross effect}}=\frac{1}{N}\sum\nolimits_{i=1}^N\left\lvert\sum\nolimits_{k\neq l}\hat{t}_{ik}\hat{a}_{il}(\boldsymbol{u}_k\cdot \boldsymbol{v}_l)\right\lvert,\\
    \overline{\text{Contrib}}_{\text{direct effect,100}}=\frac{1}{N}\sum\nolimits_{i=1}^N\left\lvert\sum\nolimits_{j=1}^{100}\hat{t}_{ij}\hat{a}_{ij}(\boldsymbol{u}_j\cdot \boldsymbol{v}_j)\right\lvert.\notag
\end{gather}
We also compute them for the negative pairs in a similar way. We obtain the mean embeddings and PLS directions on the Clotho training set, then decompose the embeddings and calculate the metrics on the Clotho testing set with these pre-calculated means and directions. In this way, generalizable results can be obtained. The results are shown in Table~\ref{tab:mean_abs_contrib_pos_neg}. 
We can see the direct effect is high for positive pairs and low for negative pairs. The cross effect is indeed relatively insignificant, which stays low for both positive pairs and negative pairs. The top 100 indices indeed account for most of the direct effect. 
Generally speaking, for many CLAP models, \textbf{the similarity is driven predominantly by the core concepts of the shared semantic head, which are compared on a one-to-one basis without considering much about the cross effects. }
\begin{table}[tbp]
    \caption{The mean abs contribution of different parts, for positive (Pos) pairs and negative (Neg) pairs separately.}
    \label{tab:mean_abs_contrib_pos_neg}
    \centering
    \begin{tabular}{lccc}
    \toprule
     Pair Type  & $\overline{\text{Contrib}}_{\text{direct effect}}$ & $\overline{\text{Contrib}}_{\text{direct effect, 100}}$ & $\overline{\text{Contrib}}_{\text{cross effect}}$ \\ \midrule
     Pos Pairs & 0.1937 & 0.1931 & 0.0326 \\
     Neg Pairs & 0.0445 & 0.0445 & 0.0226 \\\bottomrule
    \end{tabular}
\end{table}

Furthermore, for the positive pairs, the similarity $\boldsymbol{t}_i\cdot\boldsymbol{a}_i$ should be as large as possible. If a term in Eq.~(\ref{eq:simdissectfull}) produces a positive contribution, it is constructive to this target. We then count the net useful contribution produced by these terms over the entire dataset. For the cross-effect terms, we have $\sum_{i=1}^N\hat{t}_{ik}\hat{a}_{il}=\boldsymbol{u}_k^TM\boldsymbol{v}_l=(\boldsymbol{u}_k^TU)\Sigma (V^T\boldsymbol{v}_l)=0$. Thus, the net useful contribution is 0. Since we know that the cross effects are not contributing, we measure the net useful contribution of the $j$-th direction pair as: 
\begin{equation}
    \sum_{i=1}^N\hat{t}_{ij}\hat{a}_{ij}(\boldsymbol{u}_j\cdot\boldsymbol{v}_j)=(\sum_{i=1}^N\hat{t}_{ij}\hat{a}_{ij})(\boldsymbol{u}_j\cdot\boldsymbol{v}_j)=\Sigma_{jj}\times (\boldsymbol{u}_j\cdot\boldsymbol{v}_j).
\end{equation}
This justifies the use of PLS-SVD, showing why \textbf{considering both signal strength and alignment is important}. It also explains why we can \textbf{use PLS singular values $\Sigma_{jj}$ and UV alignments to find the shared semantic head: their production corresponds to the net useful contribution to the similarity of the positive pairs of the $j$-th direction~pair}. 

\begin{table*}[tbp]
\centering
\caption{Visualization of the top sentences that achieve high projection values $\hat{t}_{ij}$ on each text PLS direction $j$ from the Clotho training set. Sentences with high similarity are deduplicated before visualization. The index here begins with 0.}
\label{tab:textdirection_topsentences}
\begin{tabularx}{\textwidth}{@{} l >{\raggedright\arraybackslash}X c @{}}
\toprule
Idx & Top Sentences (projection value) & Possible Semantics \\ \midrule
0 & 
{\begin{itemize}[nosep, leftmargin=*]
    \item Heavy traffic rumbles in the background while a variety of birds are chirping and calling. (0.510)
    \item The rush of wind, traffic, and closer by birds are tweeting and chirping (0.509)
    \item A car drives away, birds chirp, followed by traffic sounds. (0.499)
    \item steady traffic travelling down a road as birds chirp in the background. (0.496)
\end{itemize}} & 
Traffic+Bird \\\midrule
1 & 
{\begin{itemize}[nosep, leftmargin=*]
    \item Vehicles are driving while men and women speak in a crowd. (0.585)
    \item numerous male and female voices in the foreground while mechanical transportation sounds spread in the background (0.574)
    \item Many men and women talk in the foreground while mechanical transportation runs. (0.553)
    \item A badly distorted speaker says something while a few customers converse. (0.545)
\end{itemize}} &
Speech\\ \midrule
2 & {\begin{itemize}[nosep, leftmargin=*]
    \item An engine that is running and rumbling in increasingly steady amounts. (0.439)
    \item The running of machinery slowly rises in pitch. (0.438)
    \item An engine runs and rumbles in increasingly steady amounts. (0.436)
    \item A motor is getting increasingly louder as it constantly runs. (0.430)
\end{itemize}} & Machinery \\ \midrule
30 & {\begin{itemize}[nosep, leftmargin=*]
    \item A coin machine was being played while coins fell loudly as time went on. (0.376)
    \item Coins are falling to the ground, with more falling as time goes on (0.354)
    \item Coins bounce and rattle as they are dropped on the cement. (0.347)
    \item Percussion plays with a voluminous reverberating echo attached. (0.341)
\end{itemize}} & \makecell[c]{Coin/\\Percussion} \\\midrule
40 & {\begin{itemize}[nosep, leftmargin=*]
    \item A siren blaring and changing in pitch in a rhythmic repeated way. (0.277)
    \item The whir of a machine escalates and then fades several times. (0.276)
    \item While a crowd of people were talking, a siren began wailing. (0.272)
    \item Large explosions intermittently as a siren continuously wails in the distance. (0.267)
\end{itemize}} & Siren \\
\bottomrule
\end{tabularx}
\end{table*}

We also visualize some directions in the shared semantic head in Table~\ref{tab:textdirection_topsentences}. We can see that \textbf{each direction roughly corresponds to interpretable concepts}. Moreover, all 1024 direction pairs can be obtained swiftly in less than 0.16s on the Clotho training set (19195 paired samples) using PyTorch on a CPU-only machine (Intel Xeon E5-2620 v4).

\section{PLSHead: Retrieval with Truncated Embeddings}\label{sec:retainhead1_retwithtruncemb}
The above insights motivate us to retain only the top 100 components (the shared head) when performing audio-text retrieval. 
We verify in this section that the performance is indeed comparable to or even better than the original, although the majority of dimensions (90\%) are discarded. 

\looseness=-1 We obtain the mean embeddings $\bar{\boldsymbol{t}},\bar{\boldsymbol{a}}$ and PLS directions $U,V$ beforehand on a training set. These parameters are then used to decompose (according to Eq.~(\ref{eq:emb_dissection})) and truncate the embeddings on a test set. Finally, the truncated test embeddings are used for evaluation. In this way, real generalization abilities can be reflected instead of observing possibly overfitted results on the training set.  

We perform experiments with both in-domain and cross-domain scenarios. We examine the in-domain performance on Clotho \cite{drossos2020clotho} and AudioCaps \cite{kim2019audiocaps}. 
These datasets are standard benchmarks for audio-text retrieval~\cite{mei2024wavcaps,bai2025audiosetcaps}. 
For the cross domain scenario, the mean and projection parameters are obtained from the Clotho training set, and the evaluation is done on the AudioCaps test set. The metrics adopted are Recall@\{1,2,10,50\}, Mean Rank, Median Rank, and mAP@10, which are consistent with prior works like \cite{mei2024wavcaps,Kim2025_t6}. The embeddings are normalized before retrieval. 

We compare the retrieval performance on these embeddings: 
\begin{enumerate}
    \item Original: the original 1024 dimensional embeddings.
    \item PLSHead: the 100 dimensional embeddings made up of the top 100 projection values $\hat{\boldsymbol{t}}_i^{1:K}=[\hat{t}_{i1},\hat{t}_{i2}\dots\hat{t}_{iK}]^T$ and $\hat{\boldsymbol{a}}_i^{1:K}=[\hat{a}_{i1},\hat{a}_{i2}\dots\hat{a}_{iK}]^T$, with $K=100$.
    \item PLSHeadWeighted (PLSHeadW): the truncated text projections $\hat{\boldsymbol{t}}_i^{1:100}$, and reweighted audio projections $\hat{\boldsymbol{a}}_i^{1:100,\text{w}}$. For the $j$-th element, $(\hat{\boldsymbol{a}}_i^{1:100,\text{w}})_j=\hat{a}_{ij}(\boldsymbol{u}_j\cdot\boldsymbol{v}_j)$.
    \item PCAHead: the 100 dimensional embeddings made up of the top 100 projection values obtained with principal component analysis (PCA). For PCA, the same setting is employed: obtaining the directions on the training set, centering the test samples with the training set means, and obtaining projections with the directions calculated from the training set. 
\end{enumerate}

\begin{table*}[tbp]
    \caption{Audio-text retrieval with truncated embeddings. The metrics are the larger the better except for MeanR/MedR.}
    \label{tab:retrieval_trunc_emb}
    \centering
    \begin{tabular*}{\textwidth}{@{\extracolsep{\fill}} l ccccccc ccccccc @{}}
    \toprule
    \multirow{2}{*}[-0.5ex]{Method} & \multicolumn{7}{@{}c}{Text to Audio} & 
\multicolumn{7}{c@{}}{Audio to Text}\\\cmidrule(lr){2-8}
\cmidrule(l){9-15}
     & 
    R1 &
    R5 &
    R10 &
    R50 &
    MeanR &
    MedR &
    mAP10 &
    R1 &
    R5 &
    R10 &
    R50 &
    MeanR &
    MedR &
    mAP10 \\\midrule
    \multicolumn{15}{c}{\textbf{In-domain, Clotho train set mean and directions $\to$ Clotho test set evaluation}}\\ \cmidrule(){1-15}
    Original & \textbf{17.42} & 39.64 & 52.19 & 80.21 & 42.36 & \underline{9} & 27.02 & \underline{21.91} & \textbf{45.45} & \textbf{58.56} & 85.26 & 30.93 & \textbf{7} & 14.25 \\
    PLSHead & \underline{17.32} & \textbf{41.21} & \underline{54.05} & \underline{82.97} & \textbf{36.30} & \underline{9} & \textbf{27.56} & \textbf{22.01} & \underline{44.88} & 57.13 & \underline{86.22} & \textbf{28.00} & \textbf{7} & \textbf{14.44} \\
    PLSHeadW & 17.30 & \underline{41.11} & \textbf{54.30} & \textbf{82.99} & \underline{36.50} & \textbf{8} & \underline{27.52} & 21.34 & 44.21 & \underline{57.22} & \textbf{86.41} & \underline{28.42} & \underline{8} & \underline{14.27} \\
    PCAHead & 0.06 & 0.42 & 0.82 & 5.05 & 504.63 & 487 & 0.23 & 0.00 & 0.29 & 0.57 & 4.31 & 1519.11 & 1163 & 0.02 \\\midrule
    \multicolumn{15}{c}{\textbf{In-domain, AudioCaps train set mean and directions $\to$ AudioCaps test set evaluation}}\\ \cmidrule(){1-15}
    Original & 28.36 & 61.13 & 75.59 & 94.21 & 13.67 & \textbf{3} & 42.25 & \textbf{40.65} & \textbf{68.86} & \underline{80.15} & \underline{96.76} & 12.22 & \textbf{2} & \textbf{26.24} \\
    PLSHead & \underline{28.97} & \underline{62.99} & \underline{76.66} & \underline{94.92} & \underline{12.36} & \textbf{3} & \underline{43.13} & 36.89 & \underline{68.34} & \textbf{80.77} & \underline{96.76} & \textbf{9.82} & \underline{3} & \underline{26.07} \\
    PLSHeadW & \textbf{29.36} & \textbf{63.39} & \textbf{76.97} & \textbf{95.07} & \textbf{12.17} & \textbf{3} & \textbf{43.49} & \underline{36.99} & 65.94 & 79.94 & \textbf{97.28} & \underline{10.19} & \underline{3} & 25.51 \\
    PCAHead & 0.36 & 1.34 & 2.49 & 12.27 & 325.93 & \underline{270} & 0.84 & 0.21 & 0.94 & 2.51 & 8.25 & 939.40 & 576 & 0.16 \\\midrule
    \multicolumn{15}{c}{\textbf{Cross-domain, Clotho train set mean and directions $\to$ AudioCaps test set evaluation}}\\ \cmidrule(){1-15}
    Original & \textbf{28.36} & \textbf{61.13} & \textbf{75.59} & \textbf{94.21} & \textbf{13.67} & \textbf{3} & \textbf{42.25} & \textbf{40.65} & \textbf{68.86} & \textbf{80.15} & \textbf{96.76} & \textbf{12.22} & \textbf{2} & \textbf{26.24} \\
    PLSHead & 27.63 & \underline{58.98} & 73.27 & \underline{93.90} & \underline{15.42} & \underline{4} & 40.90 & \underline{35.01} & 63.01 & 75.97 & \underline{95.40} & \underline{14.18} & \underline{3} & \underline{23.29} \\
    PLSHeadW & \underline{27.84} & 58.64 & \underline{73.31} & 93.79 & 15.59 & \underline{4} & \underline{41.03} & 34.80 & \underline{64.05} & \underline{76.07} & 95.30 & 14.21 & \underline{3} & 23.17 \\
    \bottomrule
\end{tabular*}
\end{table*}

\looseness=-1 As shown in Table~\ref{tab:retrieval_trunc_emb}, the PLS-based truncation methods indeed demonstrate comparable or even better performance than the original one. Note that this is achieved when the majority of the dimensions are discarded. The gain is possibly because the interferences from the noisy tail and the cross effects are removed. Our method is effective on both Clotho and AudioCaps. It can also achieve good generalization performance in cross-domain scenarios.

We see that UV alignment reweighting is not necessary for the system to work. This is possibly because the covariance of projection values already ensures that the latter dimensions contribute less. We also see that PCA cannot achieve desirable results, possibly because it is a single-modality method. In PCA, the audio and text directions are extracted separately and are not aligned.

\begin{figure}[tbp]
    \centering
    \includegraphics[width=0.87\linewidth]{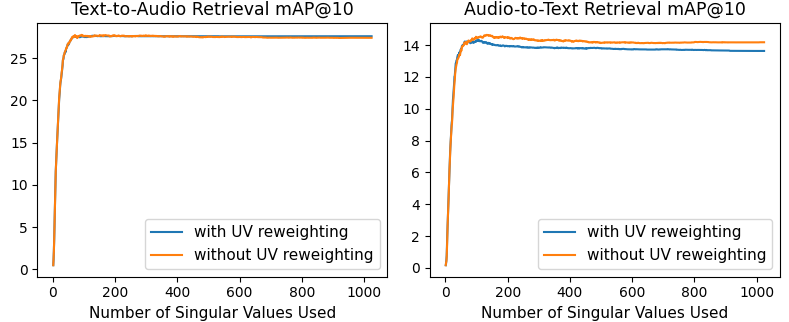}
    \caption{Retrieval performance vs number of singular values used on Clotho.}
    \label{fig:clotho_map10allthresholds}
\end{figure}

For a broader picture, we plot mAP@10 across all truncation thresholds $K$ for PLSHead and PLSHeadW in the Clotho in-domain scenario in Fig.~\ref{fig:clotho_map10allthresholds}. The performance soars within the first few indices and remains stagnant thereafter.

\looseness=-1 In practice, to compress the embeddings before transmission or storage, a device just needs a small set of precomputed mean and direction parameters. It can then process each incoming raw embedding on the run with simple operations, without needing the whole dataset or computing SVD. The stored embeddings can be directly used for retrieval, and the inner product similarity only needs to be computed on 100 dimensional embeddings instead of 1024 dimensional embeddings. This may significantly reduce the computational complexity for retrieval.

\section{Theoretical Deconstruction of Projection Decoding}\label{sec:pd_deconstruction}
\looseness=-1 Based on our decomposition framework, we proceed to investigate the modality gap. We hypothesize that, apart from the widely acknowledged mean term \cite{liang2022mind}, the unaligned tail, which retains a significant portion of the energy (Sec.~\ref{sec:dissect_clap_emb}), could also be a key source of the gap. 
To validate this hypothesis, we apply our framework to explain Projection Decoding (PD). While PD has emerged as a highly effective post-hoc method for gap reduction \cite{li2023decap,Kouzelis2023,li2025drcap}, its underlying mechanisms have previously lacked theoretical explanation. By analyzing PD through our framework, we provide this missing explanation and confirm our hypothesis, which naturally lays the foundation for our direct approach proposed in the next section.

To make the discussion concrete, we focus on the setting of zero-shot audio captioning. To map an audio embedding $\boldsymbol{a}$ into the text space, PD \cite{li2023decap} uses a large memory of text embeddings $X\in\mathbb{R}^{N\times C}$ (usually formed by stacking all training set embeddings row-wise\cite{li2023decap,Kouzelis2023,li2025drcap}, here denoted as training text embeddings $\boldsymbol{t}_i$). The new embedding $\boldsymbol{t}^{\text{pd}(\boldsymbol{a})}$ is a linear combination of these memory embeddings $\boldsymbol{t}_i$, weighted by their similarity to the audio embedding $\boldsymbol{t}_i\cdot\boldsymbol{a}$:
\begin{equation}\label{eq:pd}
    \boldsymbol{t}^{\text{pd}(\boldsymbol{a})}= \sum_{i=1}^{N}\frac{\exp\bigl[(\boldsymbol{t}_i\cdot\boldsymbol{a})/\tau\bigr]}{\sum_{j=1}^N\exp\bigl[(\boldsymbol{t}_j\cdot\boldsymbol{a})/\tau\bigr]}\boldsymbol{t}_i,
\end{equation}
where $\tau$ is a small temperature parameter. Here, softmax is employed to normalize the similarity weights, which is a nonlinear operation. To build intuition, we first analyze a simplified linear version of PD. Then, we discuss the behavior of the actual PD and provide experimental verifications.
\subsection{Linear PD}
\begin{figure}
\centering
\subfloat[The full matrix]{
    \includegraphics[width=0.47\columnwidth]{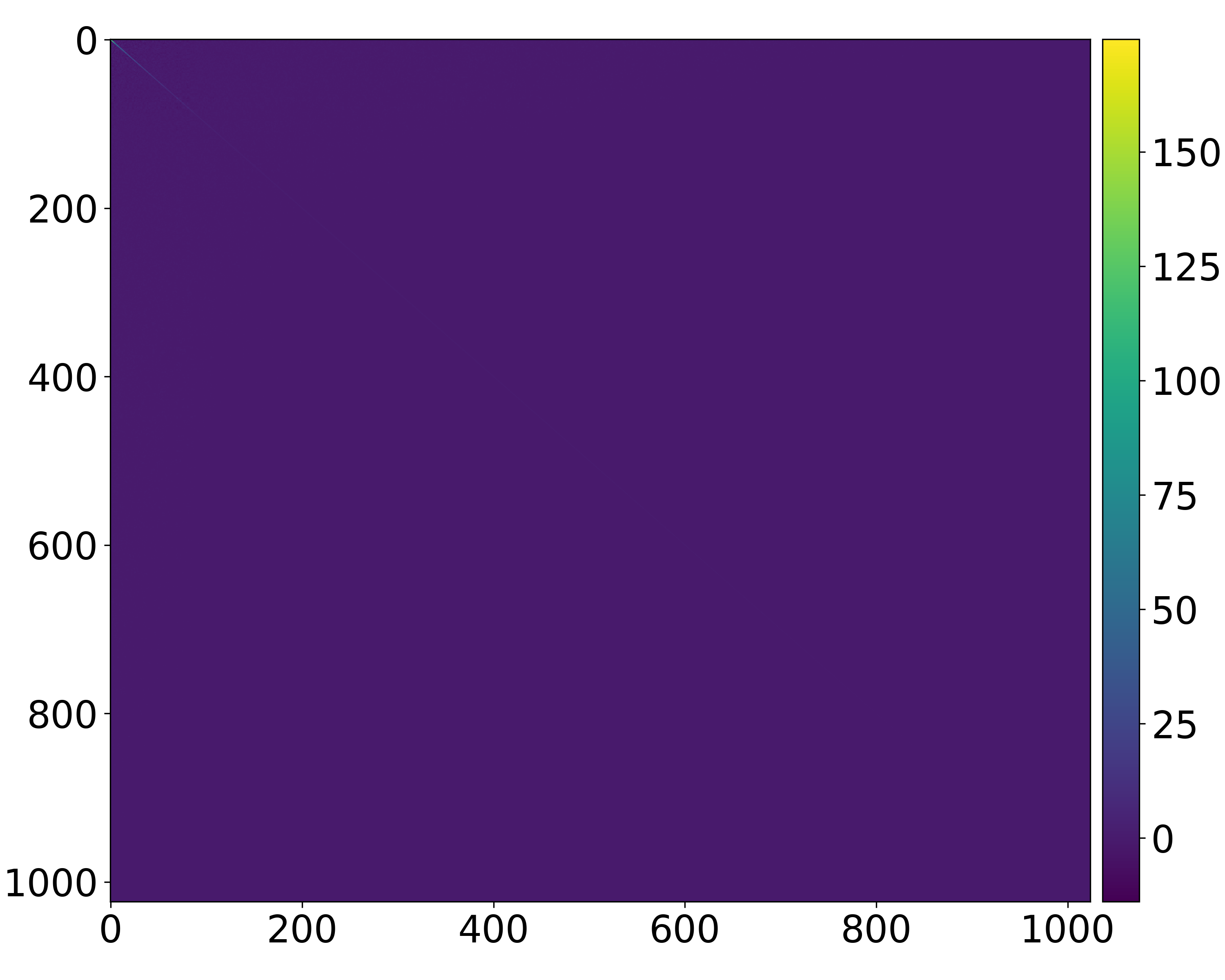}
}
\subfloat[Zoom-in of first 250$\times$250 elements]{
    \includegraphics[width=0.47\columnwidth]{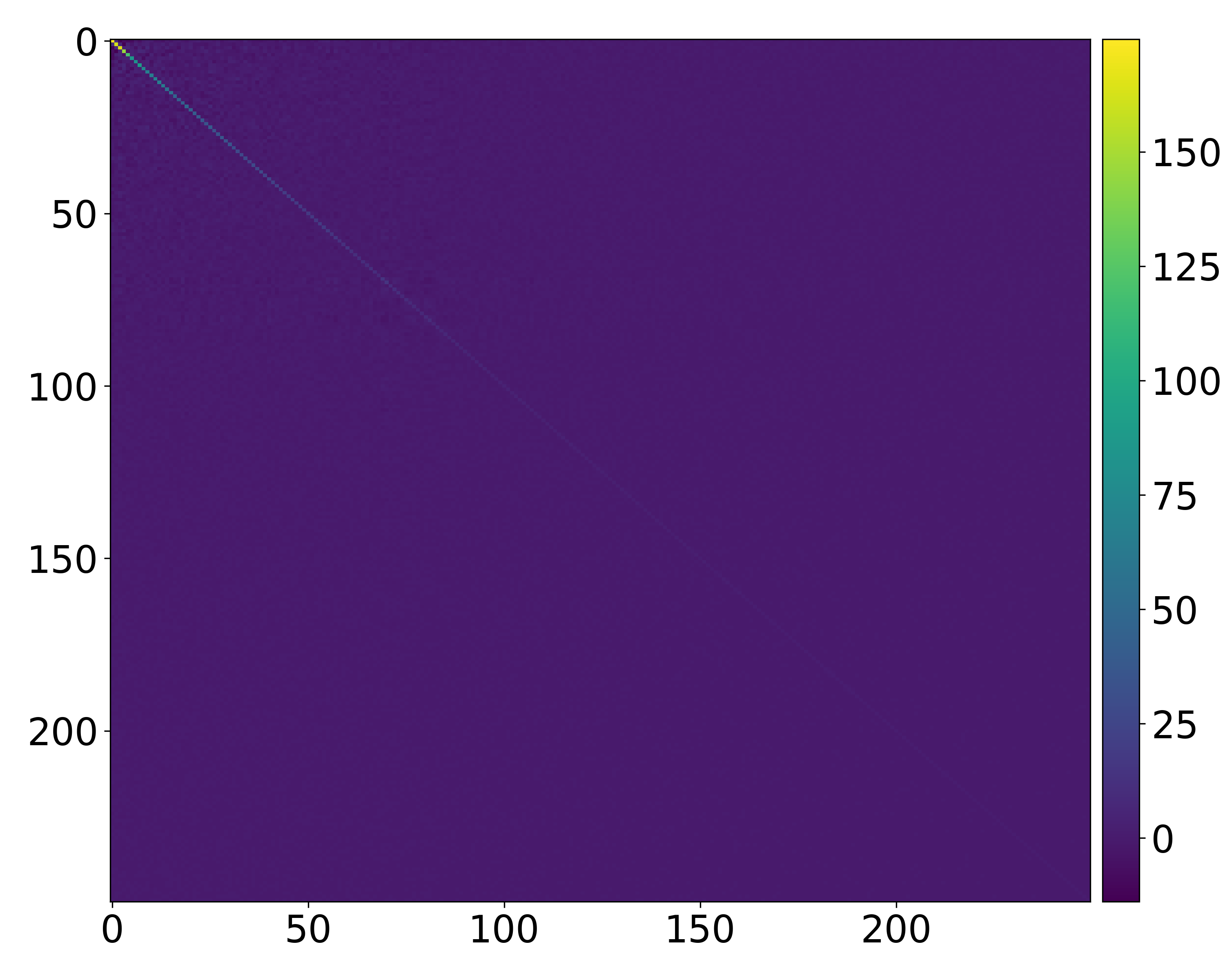}
}
    \caption{\textbf{Visualization of $\hat{X}^T\hat{X}$.} Note the upper-left diagonal area.} 
    \label{fig:xprojselfcov}
\end{figure}

To simplify the original PD into a linear version, we directly use the similarities $\boldsymbol{t}_i\cdot\boldsymbol{a}$ without going through the softmax. The scale of the output $\boldsymbol{t}^{\text{pd}(\boldsymbol{a})}$ can be normalized to one afterwards. For simplicity, we assume that the audio embedding $\boldsymbol{a}$ and text memory embeddings $\boldsymbol{t}_i$ do not have a mean component here. The linear PD is formulated as: 
\begin{equation}\label{eq:linearpd}
    \boldsymbol{t}^{\text{pd}(\boldsymbol{a})}=\sum_{i=1}^N(\boldsymbol{t}_i\cdot\boldsymbol{a})\boldsymbol{t}_i=X^T(X\boldsymbol{a}).
\end{equation}
We decompose with $X=\hat{X}U^T$ and $\boldsymbol{a}=V\hat{\boldsymbol{a}}$. Here, $U,V\in\mathbb{R}^{C\times C}$ are the (column-wise stacked) text and audio PLS concept directions. $\hat{\boldsymbol{a}}\in\mathbb{R}^C$ stores the projection coefficients of $\boldsymbol{a}$ on $V$, while $\hat{X}\in\mathbb{R}^{N\times C}$ stores the (row-wise stacked) projection coefficients of all $\boldsymbol{t}_i$ on $U$. We have:
\begin{align}\label{eq:linear_pd_dissection}
    \boldsymbol{t}^{\text{pd}(\boldsymbol{a})}&=X^T(Xa)=(\hat{X}U^T)^T(\hat{X}U^TV\hat{\boldsymbol{a}})\notag\\
    &=U\underbrace{(\hat{X}^T\hat{X})}_{\text{Rescaling}}\underbrace{(U^TV)}_{\text{Filtering}}\hat{\boldsymbol{a}}.
\end{align}
\looseness=-1 Note that $U^TV$ is the UV matrix in Fig.~\ref{fig:uvmatrix}. It is near-zero except for the $\sim$top 100 (shared head) diagonal values. These top diagonal values are large and positive. $\hat{X}^T\hat{X}$ is the covariance of the text projection coefficients in the training data. Its diagonals are the self-variances (see Fig.~\ref{fig:clotho_sdecomp}). $\hat{X}^T\hat{X}$ is shown in Fig.~\ref{fig:xprojselfcov}. We see a similar pattern: near-zero off-diagonal entries, along with quickly decaying diagonal values as the index increases. 

Eq.~(\ref{eq:linear_pd_dissection}) reveals that Linear PD is a sequence of three operations applied to the audio projection coefficients $\hat{\boldsymbol{a}}$: \textbf{1)} Filtering out the noisy, unaligned private tail ($\sim$last 924 dim) with the alignment gate $U^TV$; \textbf{2)} Rescaling the filtered coefficients according to the self-covariance $\hat{X}^T\hat{X}$. It further promotes the head and demotes the tail; \textbf{3)} Synthesizing $\boldsymbol{t}^{\text{pd}(\boldsymbol{a})}$ with the text directions $U$. This swaps the original audio directions $V$ with the new text directions.

\subsection{Actual PD}
The linear PD confirms our intuition about the modality gap. 
The actual PD additionally applies softmax to the similarity scores. We now show that the behavior can still be similar. 

\looseness=-1 In fact, the softmax temperature $\tau$ used in PD is a tiny value like 0.01 \cite{li2023decap,Kouzelis2023}, greatly sharpening the similarity scores. 
Also, we have known in Sec.~\ref{sec:retainhead1_retwithtruncemb} that the retrieval approximately just compares projections in the top PLS directions. 
Thus, the softmax virtually preserves only a very small number of memory embeddings that share a similar head with the input audio embedding. This indicates that for a preserved text embedding $\boldsymbol{t}_{i}^{\text{sel}}$ from the memory, its projection on the text directions, which is $\hat{\boldsymbol{t}}_{i}^{\text{sel}}$, is similar to the projection of $\boldsymbol{a}$ on the audio directions, which is $\hat{\boldsymbol{a}}$, for the top projection indices $j=1\dots K$. We denote this as $(\hat{\boldsymbol{t}}_{i}^{\text{sel}})^{1:K}\approx (\hat{\boldsymbol{a}})^{1:K}$, where $(\cdot)^{1:K}\in\mathbb{R}^K$ means selecting indices from 1 to $K$ to form a sub-vector.

\looseness=-1 The PD, which is a weighted sum of these selected items $\boldsymbol{t}_{i}^{\text{sel}}$, will now produce an output embedding $\boldsymbol{t}^{\text{pd}(\boldsymbol{a})}$ that has a head similar to the original. Since it is averaged from multiple similar heads, noise could be reduced more compared to nearest-neighbor decoding (NND), which uses only one top-similarity sample. Thus, the similarity to the original head $(\hat{\boldsymbol{a}})^{1:K}$ should be higher for the PD head than the NND head. This indicates that the projection of $\boldsymbol{t}^{\text{pd}(\boldsymbol{a})}$ onto the text directions, which is $\widehat{\boldsymbol{t}^{\text{pd}(\boldsymbol{a})}}$, is similar to the projection of the original embedding $\boldsymbol{a}$ onto the audio directions, which is $\hat{\boldsymbol{a}}$, for the top indices 1 to $K$: $(\widehat{\boldsymbol{t}^{\text{pd}(\boldsymbol{a})}})^{1:K}\approx (\hat{\boldsymbol{a}})^{1:K}$.

For the tail, we can easily see that the original one $(\hat{\boldsymbol{a}})^{K+1:C}$ is now stripped, and a new tail is now constructed as the weighted sum of the selected memory embeddings. This means that although the new embedding $\boldsymbol{t}^{\text{pd}(\boldsymbol{a})}$ still has a tail, it could be now drastically different from the original one.

Moreover, for the mean term, since the new embedding $\boldsymbol{t}^{\text{pd}(\boldsymbol{a})}$ is constructed from averaging the text embeddings $\boldsymbol{t}_{i}^{\text{sel}}$, and each text embedding has a stable mean component $\bar{\boldsymbol{t}}$, the averaged embedding should still preserve this item. Thus, it could also be effectively doing the work of ``mean shifting'', which is swapping out the original mean term $\bar{\boldsymbol{a}}$, and adding the new one $\bar{\boldsymbol{t}}$ to the output $\boldsymbol{t}^{\text{pd}(\boldsymbol{a})}$.

To summarize, the projection decoding can be viewed as effectively performing\footnote{A mathematical derivation is in App.~\ref{sec:app_derive_actualpd}, which corroborates the view.}:
\begin{enumerate}
    \item \textbf{Head preserving.} Get the projection values of the original embedding, preserve the head, and discard the tail.
    \item \textbf{Tail imputation.} Find embeddings in the memory that share a similar head, and build a new tail as a weighted average of the tails of these selected embeddings.
    \item \textbf{Base changing.} Change from $V$ to $U$ when reconstructing the embeddings from projection values.
    \item \textbf{Mean shifting.} Strip the original mean term of the source modality (audio mean $\bar{\boldsymbol{a}}$), and add the mean term of the target modality (text mean $\bar{\boldsymbol{t}}$).
\end{enumerate}

\begin{table}[tbp]
    \caption{Characterization of Projection Decoding Behaviors.}
    \label{tab:pd_exp}
    \centering
    {
    \setlength{\aboverulesep}{0pt}
    \setlength{\belowrulesep}{0pt}
    \renewcommand{\arraystretch}{1.4}
    \setlength\tabcolsep{3pt}
    \begin{tabular}{lc||lc}\toprule
        Metric & Value & Metric & Value \\\midrule
        $\operatorname{sim_{cos}}(\bar{\boldsymbol{a}}^{\text{(test)}},\bar{\boldsymbol{t}})$ & 0.598 & $\|\bar{\boldsymbol{a}}^{\text{(test)}}-\bar{\boldsymbol{t}}\|$ & 0.488 \\
       $\operatorname{sim_{cos}}(\overline{\boldsymbol{t}^{\text{pd}(\boldsymbol{a})}},\bar{\boldsymbol{t}})$  & 0.995 & $\|\overline{\boldsymbol{t}^{\text{pd}(\boldsymbol{a})}}-\bar{\boldsymbol{t}}\|$ & 0.112 \\
       $\overline{\operatorname{sim_{cos}}\bigl((\widehat{\boldsymbol{t}^{\text{pd}(\boldsymbol{a})}})^{1:100},(\hat{\boldsymbol{a}})^{1:100}\bigr)}$ & 0.772 & $\operatorname{sim_{cos}}(\bar{\boldsymbol{a}}^{\text{(train)}},\bar{\boldsymbol{a}}^{\text{(test)}})$ & 0.999 \\
       $\overline{\operatorname{sim_{cos}}\bigl((\widehat{\boldsymbol{t}^{\text{pd}(\boldsymbol{a})}})^{101:C},(\hat{\boldsymbol{a}})^{101:C}\bigr)}$ & 0.095 & $\|\bar{\boldsymbol{a}}^{\text{(train)}}-\bar{\boldsymbol{a}}^{\text{(test)}}\|$ & 0.028 \\
      $\overline{\operatorname{sim_{cos}}\bigl((\widehat{\boldsymbol{t}^{\text{nnd}(\boldsymbol{a})}})^{1:100},(\hat{\boldsymbol{a}})^{1:100}\bigr)}$ & 0.677 &  &  \\\bottomrule
    \end{tabular}
    }
\end{table}
We also performed experiments to verify our theory, with the results given in Table~\ref{tab:pd_exp}. We perform PD on all Clotho test set audio embeddings according to Eq.~(\ref{eq:pd}), using the Clotho training set text embeddings as the memory bank. These embeddings are further normalized to an $l2$-norm of 1, which is consistent with previous implementations \cite{li2023decap,Kouzelis2023}. The resultant embeddings are $\boldsymbol{t}^{\text{pd}(\boldsymbol{a}_i)},i=1\dots N_{\text{test}}$.

We first see whether the mean term is shifted from $\bar{\boldsymbol{a}}$ to $\bar{\boldsymbol{t}}$. We calculate the mean embedding of all PD embeddings $\overline{\boldsymbol{t}^{\text{pd}(\boldsymbol{a})}}=(\sum_{i=1}^{N_{\text{test}}}\boldsymbol{t}^{\text{pd}(\boldsymbol{a}_i)})/N_{\text{test}}$. Then, its cosine similarity to the target text mean $\operatorname{sim_{cos}}(\overline{\boldsymbol{t}^{\text{pd}(\boldsymbol{a})}},\bar{\boldsymbol{t}})$ is calculated, where $\bar{\boldsymbol{t}}$ is the mean embedding of all the text embeddings in the training set. It is compared with the similarity before PD mapping $\operatorname{sim_{cos}}(\bar{\boldsymbol{a}}^{\text{(test)}},\bar{\boldsymbol{t}})$, where $\bar{\boldsymbol{a}}^{\text{(test)}}$ is the mean embedding calculated on all audio embeddings in the test set\footnote{Actually, the mean terms are relatively stable. We can see that audio means from the test set $\bar{\boldsymbol{a}}^{\text{(test)}}$ and training set $\bar{\boldsymbol{a}}^{\text{(train)}}$ are very close. This is consistent with the findings in other works \cite{liang2022mind,zhang2024connect}.They are not distinguished later.}. We see that after PD, the mean $\overline{\boldsymbol{t}^{\text{pd}(\boldsymbol{a})}}$ is now very similar to the text mean $\bar{\boldsymbol{t}}$, which is much better than the original one before PD mapping $\operatorname{sim_{cos}}(\bar{\boldsymbol{a}}^{\text{(test)}},\bar{\boldsymbol{t}})$. We also calculated the distance between the means before and after PD mapping $\|\bar{\boldsymbol{a}}^{\text{(test)}}-\bar{\boldsymbol{t}}\|$ and $\|\overline{\boldsymbol{t}^{\text{pd}(\boldsymbol{a})}}-\bar{\boldsymbol{t}}\|$. We see that after PD, the distance is significantly reduced. This indicates that mean shifting is indeed performed in PD.

Knowing that the mean term in the new PD embedding is basically the text mean $\bar{\boldsymbol{t}}$, we proceed to the decomposition analysis of the head and tail using the PLS decomposition in Eq.~(\ref{eq:emb_dissection}). Here, the decomposition parameters $U,V,\bar{\boldsymbol{t}},\bar{\boldsymbol{a}}$ are derived from the training set. The PD-mapped embeddings $\boldsymbol{t}^{\text{pd}(\boldsymbol{a}_i)}$ are decomposed with $U,\bar{\boldsymbol{t}}$. The raw testing audio embeddings are decomposed with $V,\bar{\boldsymbol{a}}$. We see that the average (over all test samples $\boldsymbol{a}_i$,$i=1\dots N_{\text{test}}$) cosine similarity of the head before and after PD-mapping is considerably high (0.772), and is higher than that of nearest-neighbor decoding (denoted as $\boldsymbol{t}^{\text{nnd}(\boldsymbol{a}_i)}$, 0.677). We also see the tail is indeed different before and after PD mapping (0.095).

\looseness=-1 Based on these insights, we deduce that the modality gap may have multiple sources that reside in different parts of the embedding space. We categorize these possible sources according to the subspace as follows:
\begin{enumerate}
    \item \textbf{Mean Components.} The difference between $\bar{\boldsymbol{t}}$ and $\bar{\boldsymbol{a}}$ is the static modality gap in mainstream works \cite{liang2022mind,zhang2023diagnosing,shi2023towards,zhang2024connect,li2025closing,yaras2025explaining,grassucci2025closing}. $\bar{\boldsymbol{t}}$ and $\bar{\boldsymbol{a}}$ are stable across samples. It is usually explained with the cone effect\cite{liang2022mind} introduced in Sec.~\ref{sec:rel_work_mod_gap}. It can be reduced with embedding shift (ES). 
    \item \textbf{Shared Head.} \looseness=-1 The PLS directions $\boldsymbol{u}_i$ and $\boldsymbol{v}_i$ are not perfectly aligned, which could be a source of the gap. Shifting to the same set of directions may be needed to address it (effects shown later in Table~\ref{tab:audiocaptioningeval}). 
    Also, differences in the estimated semantics in the audio and text projection coefficients could be another source of the gap.
    \item \textbf{Modality-specific Tail.} The entire tail can be a gap to be removed. The tails are unaligned and have significant residual energy. To mitigate this, such as mapping an audio embedding into the text space, we can strip the original audio tail and ``hallucinate'' a text tail based on some similar text embeddings in a memory. Or, we can remove all tails in text and audio embeddings entirely.
\end{enumerate}

\section{PLSHead II: Bridging the Modality Gap for Audio Captioning}
\begin{table}[tbp]
    \centering
    \caption{Results on Audio captioning.}
    \label{tab:audiocaptioningeval}
    \setlength\tabcolsep{3.6pt}
    \begin{threeparttable}
    \begin{tabular}{lcccccc}\toprule
        Method & BLEU$_4^{\uparrow}$ & MET$^{\uparrow}$ & R$_{\text{L}}^{\uparrow}$ & CIDEr$^{\uparrow}$ & SPICE$^{\uparrow}$ & SPIDEr$^{\uparrow}$ \\\midrule
        \multicolumn{7}{c}{\textbf{Clotho, CLAP: WavCaps HTSAT-BERT-ZS\cite{mei2024wavcaps}}}\\ \cmidrule(){1-7}
        t$\to$a AD* & 
        11.1&	14.7&	33.3&	26.6&	8.6&	17.6\\
        t$\to$a NI*&
        12.5&	15.5&	33.5&	27.4&	10.3&	18.8\\
        t$\to$a ES* & 
        11.6&	16.3&	34.3&	30.8&	10.6&	20.7\\
        t$\to$a NND* & 
        13.1&	17.2&	35.7&	36.0&	12.3&	24.1\\
        t$\to$a PD* & 
        15.1&	\underline{17.9}&	\underline{37.5}&	\textbf{42.3}&	\underline{13.0}&	\textbf{27.7}\\
        t\textsuperscript{100}$\to$a\textsuperscript{100}$\star$* & 
        \underline{15.4}&\textbf{18.3}&\textbf{37.8}&\underline{41.8}&\textbf{13.3}&\underline{27.5} \\
        t\textsuperscript{100rec}$\to$a\textsuperscript{100rec}*&
        14.1&17.4&36.4&36.0&12.1&24.0 \\
        WSAC \cite{Kouzelis2023}$\dagger$&
        12.6&16.9&35.9&35.7&11.8&23.8 \\
        SoftHard\cite{zhang2024zero} & 
        \textbf{15.6}&17.3&\underline{37.5}&40.3&11.9&26.1\\\midrule
        a$\to$a*&\textbf{16.6}&\textbf{18.0}&\textbf{38.0}&41.1&\textbf{12.6}&26.8 \\
        a\textsuperscript{100}$\to$a\textsuperscript{100}$\star$*&16.5&17.9&37.6&\textbf{42.0}&12.4&\textbf{27.2}\\\midrule
        t\textsuperscript{-924}$\to$t\textsuperscript{-924}*& 7.1&11.4&27.1&12.0&5.7&8.8\\
        a\textsuperscript{-924}$\to$a\textsuperscript{-924}*& 6.2&11.0&27.2&7.5&5.0&6.3\\
        t\textsuperscript{-924}$\to$a\textsuperscript{-924}*& 5.9&10.9&26.0&8.8&4.9&6.9\\
        no\_cond*& 3.5&8.5&24.9&5.4&2.3&3.9\\\midrule
        \multicolumn{7}{c}{\textbf{AudioCaps, CLAP: WavCaps HTSAT-BERT-ZS\cite{mei2024wavcaps}}}\\\cmidrule(){1-7}
        t$\to$a AD* & 
        14.6&	18.5&	36.2&	34.4&	12.2&	23.3\\
        t$\to$a ES* &
        16.0&	20.9&	40.8&	46.8&	13.9&	30.4\\
        t$\to$a NND* &
        21.0&	23.8&	44.1&	57.9&	\underline{17.1}&	37.5\\
        t$\to$a PD* &
        \textbf{23.0}&	\textbf{24.9}&	\textbf{47.1}&	\textbf{65.1}&	\textbf{17.8}&	\textbf{41.5}\\
        t\textsuperscript{100}$\to$a\textsuperscript{100}$\star$* & 
        \underline{21.9}&	\underline{24.2}&	\underline{46.7}&	64.1&	\underline{17.1}&	\underline{40.6}\\
        t\textsuperscript{100rec}$\to$a\textsuperscript{100rec}*&
        16.7&21.7&42.1&49.7&14.8&32.2 \\
        WSAC \cite{Kouzelis2023}$\dagger$ & 
        17.1&	23.2&	43.5&	56.4&	16.3&	36.3\\
        SoftHard\cite{zhang2024zero} & 
        21.3&	22.0&	45.7&	\underline{64.4}&	15.6&	40.0\\\midrule
        a$\to$a* & 
        \textbf{23.7}&	\textbf{23.8}&	\textbf{47.3}&	\textbf{63.0}&	\textbf{17.7}&	\textbf{40.3}\\
        a\textsuperscript{100}$\to$a\textsuperscript{100}$\star$* & 
        21.7&	23.7&	46.8&	60.7&	17.1&	38.9\\\midrule
        \multicolumn{7}{c}{\textbf{Clotho, CLAP: WavCaps+SoundVECaps as in DRCap\cite{li2025drcap}}}\\ \cmidrule(){1-7}
        t$\to$a AD* & 9.0&14.6&30.7&24.0&8.6&16.3\\
        t$\to$a ES* & 12.3&16.6&34.9&32.5&11.2&21.9\\
        t$\to$a PD* & \textbf{15.6}&18.1&\textbf{37.9}&\underline{43.1}&\underline{13.2}&\underline{28.2}\\
        t\textsuperscript{100}$\to$a\textsuperscript{100}$\star$*&\underline{15.3}&\textbf{18.5}&\underline{37.7}&41.6&\textbf{13.3}&27.4\\
        DRCap\cite{li2025drcap}&-&\underline{18.2}&-&\textbf{43.8}&\textbf{13.3}&\textbf{28.5}\\\midrule
        a$\to$a* & \textbf{16.3}&\textbf{17.7}&\textbf{37.9}&\textbf{41.6}&\textbf{12.4}&\textbf{27.0}\\
        a\textsuperscript{100}$\to$a\textsuperscript{100}$\star$* & \textbf{16.3}&\textbf{17.7}&37.5&41.0&12.3&26.7\\\midrule
        \multicolumn{7}{c}{\textbf{AudioCaps, CLAP: WavCaps+SoundVECaps as in DRCap\cite{li2025drcap}}}\\ \cmidrule(){1-7}
        t$\to$a AD* & 
        20.4&	23.4&	43.4&	51.1&	16.2&	33.7\\
        t$\to$a ES* & 
        21.8&	22.8&	45.8&	58.6&	16.3&	37.5\\
        t$\to$a PD* & 
        \textbf{25.9}&	\underline{24.9}&	\textbf{49.9}&	\textbf{70.5}&	\textbf{18.5}&	\textbf{44.5}\\
        t\textsuperscript{100}$\to$a\textsuperscript{100}$\star$*&
        \underline{25.1}&	24.6&	\underline{49.7}&	\underline{69.2}&	17.5&	43.3\\
        DRCap\cite{li2025drcap}&
        -&		\textbf{25.3}&	-&		\textbf{70.5}&	\underline{18.0}&	\underline{44.2}\\\midrule
        a$\to$a* & 
        \textbf{25.3}&	\textbf{24.1}&	\textbf{48.7}&	\textbf{66.6}&	\textbf{18.2}&	\textbf{42.4}\\
        a\textsuperscript{100}$\to$a\textsuperscript{100}$\star$* & 
        23.4&	23.7&	47.8&	62.8&	17.7&	40.2\\\bottomrule
    \end{tabular}
    \begin{tablenotes}
        \footnotesize
        \item[*] results produced by us based on modified WSAC \cite{Kouzelis2023} that uses beam\_5. 
        \item[$\star$] 100 dim embeddings as opposed to the original 1024 dim ones.
        \item[$\dagger$] results provided in Zhang et al. \cite{zhang2024zero} that use the same CLAP model.
    \end{tablenotes}
    \end{threeparttable}
\end{table}

From the observations above, we find that the embedding of top $K$ ($K$ is the head size, which is 100 here) projection values is a promising solution that not only retains most of the useful information but also avoids most of the pitfalls of the modality gap mentioned above. 
In this section, we study whether these truncated projection embeddings can both reduce the dimensionality from 1024 to 100, and bridge the modality gap for zero-shot audio captioning. We also look at its performance in fully-supervised audio captioning.

\looseness=-1 We base our experiments on WSAC \cite{Kouzelis2023}, which is an early method for zero-shot audio captioning. It adopts a simple model (Transformer with 4 layers, 4 heads, and 768 dim hidden size, not pretrained), and a simple pipeline that purely utilizes condition swapping and basic gap mitigation methods (NI, ES, NND, or PD). Later works also add more sophisticated modules, such as extra keyword guidance and advanced data augmentation \cite{zhang2024zero}, or extra similar caption guidance based on a massive datastore of 450k captions\cite{li2025drcap}. They also utilize a larger captioning model, such as pretrained GPT-2 \cite{radford2019language,zhang2024zero}, LoRA \cite{hu2022lora} finetuned Vicuna-7b-v1.5 \cite{zheng2023judging} (Llama 2 \cite{touvron2023llama} finetuned on chat data)\cite{li2025drcap}. Nevertheless, we show that the performance of this simple method can approach these more sophisticated methods with some simple parameter changes, such as replacing the greedy search (hardly used in newer works) in the original WSAC \cite{Kouzelis2023} to beam decoding. Specifically, we use beam decoding with a beam size of 5, repetition penalty of 1.2, n-gram repetition limit of 3, and MBR \cite{kumar2004minimum} selection among the 5 beam results (select a caption from the 5 beam results that has the highest BLEU\textsubscript{4} similarity to the other 4 results). The epoch is 30 on Clotho and 5 on AudioCaps, due to different numbers of unique samples in the datasets. Audio clips are cropped or padded to 10 sec when HTSAT \cite{chen2022hts} encoder is used, like in \cite{zhang2024zero}. Other parameters mainly follow the original WSAC paper \cite{Kouzelis2023}. These settings are kept the same for all results produced by us in Table~\ref{tab:audiocaptioningeval}. We perform experiments on Clotho \cite{drossos2020clotho} and AudioCaps \cite{kim2019audiocaps}. The projection parameters are obtained on the training data. We show results on two CLAP models: 1) HTSAT-BERT-ZS pretrained on WavCaps \cite{mei2024wavcaps}, which is used in SoftHard \cite{zhang2024zero}, and 2) the CLAP model used in DRCap \cite{li2025drcap}, which is pretrained on WavCaps \cite{mei2024wavcaps} and SoundVECaps \cite{yuan2025sound}. In this way, the results can be compared with these reference models. Also, both HTSAT-BERT-ZS \cite{mei2024wavcaps} and the CLAP in DRCap \cite{li2025drcap} exclude overlapping samples from Clotho and AudioCaps during pretraining, which better evaluates zero-shot and generalization abilities. The metrics are BLEU\textsubscript{4} \cite{papineni2002bleu}, METEOR (MET) \cite{banerjee2005meteor}, ROUGE\textsubscript{L} (R\textsubscript{L}) \cite{lin2004rouge}, CIDEr \cite{vedantam2015cider}, SPICE \cite{anderson2016spice}, and SPIDEr \cite{liu2017improved}, which are standard audio captioning metrics \cite{mei2024wavcaps,bai2025audiosetcaps,Kouzelis2023,li2025drcap,zhang2024zero}, and reflect the correspondence to ground-truth captions. Higher values mean better performance.

\looseness=-1 The results are shown in Table~\ref{tab:audiocaptioningeval}. We show results of text-only training of our modified WSAC using the original CLAP embeddings as ``t$\to$a'' (training-time condition$\to$test-time condition). ``AD'' means no additional gap mitigation methods are used. ``NI'' means Gaussian noise injection with variance 0.013 (recommended value in WSAC \cite{Kouzelis2023}), followed by normalization, is used during training. ``ES'' means embedding shift that subtracts the static gap vector $\bar{\boldsymbol{a}}-\bar{\boldsymbol{t}}$ from the audio embeddings, followed by normalization, during inference. ``NND'' is nearest-neighbor decoding. It selects the text embedding (from a memory of all training text embeddings) that is most similar to the input audio embedding as the condition given to the captioning model. ``PD'' is projection decoding, which also utilizes a memory of all training text embeddings. We see that the performance is generally AD$<$NI$<$ES$<$NND$<$PD. The ``AD'' performs much worse than other methods, indicating the necessity of gap mitigation methods. After the simple changes, the performance of PD now increases sharply over the original greedy-decoding based results, and even generally matches or surpasses that of more sophisticated methods like SoftHard \cite{zhang2024zero} and DRCap \cite{li2025drcap}. This demonstrates the power of PD. However, it depends on large memory banks, and operates on original embedding dimensions (1024 here). Our PLSHead method, which uses top 100 projection values for text embeddings (t\textsuperscript{100}) at training time, and top 100 audio projection values (a\textsuperscript{100}) at test time, generally has a performance matching the PD results. Also, our method reduces the embedding dimension over 90\%, and eliminates the need of a large, non-parametric memory bank. Here, it simply needs 202 embeddings for mean and projection directions. 

Furthermore, we show results of fully-supervised audio captioning as a$\to$a. We see that the zero-shot PD results, zero-shot t\textsuperscript{100}$\to$a\textsuperscript{100} results, and fully-supervised a\textsuperscript{100}$\to$a\textsuperscript{100} results generally match the performance of a$\to$a. This indicates that both PD and PLSHead can push the performance of zero-shot audio captioning near the boundary of its fully-supervised counterpart. Also, compressed audio embeddings with PLSHead has matching performance with the uncompressed ones in audio captioning. These results further corroborate the view that most of the useful semantics can reside within these compact head dimensions, and these semantics are shared between embeddings of different modalities. 

Moreover, to see whether shifting the projection directions $U$ and $V$ in the head to the same set of orthogonal directions is helpful for gap mitigation, we show results of t\textsuperscript{100rec}$\to$a\textsuperscript{100rec}. It reconstructs the 1024 dim audio embeddings from the top 100 audio projection values $\hat{\boldsymbol{a}}$ (pad the remaining dimensions 101 to 1024 with 0 for ease of annotation here) and the audio directions $V$ as $V\hat{\boldsymbol{a}}$. The text embeddings are reconstructed in a similar way. We see that the performance is indeed lower than t\textsuperscript{100}$\to$a\textsuperscript{100}. This indicates that the not fully aligned $U$ and $V$, although might function in retrieval, could serve as a source of the modality gap that makes a captioner difficult to generalize between the embeddings of different modalities.

To further understand what can be in the tail dimensions, we perform additional experiments using the last 924 projection coefficients (all coefficients excluding the 100 dim head), denoted as t\textsuperscript{-924} for text tails, and a\textsuperscript{-924} for audio tails. We also perform experiments that use no condition at all, which is denoted as ``no\_cond''. 
We see that training with text tails and testing with audio tails (t\textsuperscript{-924}$\to$a\textsuperscript{-924}) yields a very low performance, which is only a little higher than the ``no\_cond'' results. This indicates that there is not much aligned information for caption prediction left in the tail dimensions. The result of a\textsuperscript{-924}$\to$a\textsuperscript{-924} is similar to t\textsuperscript{-924}$\to$a\textsuperscript{-924}, indicating that there is not much additional information for caption prediction available in the modality-private subspace of the audio embeddings other than the little amount of residual aligned information mentioned in t\textsuperscript{-924}$\to$a\textsuperscript{-924}. For t\textsuperscript{-924}$\to$t\textsuperscript{-924}, the performance is a little higher, but not substantial. This suggests that the tail of the text modality may contain additional text-specific information that provides slight benefits for caption prediction.

\section{Conclusion}
\looseness=-1 We have developed a systematic framework for understanding CLAP embeddings with PLS-SVD. We find that the CLAP embedding space can be decomposed into a mean component, which encodes the static modality gap found in other works; a compact head of shared core semantics; and a long, unaligned tail that contains significant residual energy. The modality gap could also reside within the imperfect UV alignments in the head, and importantly, the unaligned tail. The directions in the head roughly correspond to interpretable concepts. We find that only the direct effect of the shared head is actively participating in the similarity calculation. In addition, the singular values of PLS-SVD and the UV alignments can characterize the net useful contribution in aligning the positive pairs. Moreover, we can preserve just the compact head (named PLSHead) to strongly compress embeddings, and yield comparable or better performance on audio-text retrieval and audio captioning. Starting with a linear approximation, we find that projection decoding shifts the mean, preserves the head, and constructs a different tail. We find that PLSHead also effectively mitigates the modality gap in text-only training for audio captioning, and that there is not much caption-related information left in the tails, which corroborates our theory.

\appendices
\section{Theoretical Derivation of the Actual PD}\label{sec:app_derive_actualpd}
We also provide a theoretical derivation of the actual PD here. Starting from Eq.~(\ref{eq:pd}), we have:
\begin{equation}\label{eq:actualpd_1}
    \begin{aligned}
        \boldsymbol{t}^{\text{pd}(\boldsymbol{a})}&=X^T\operatorname{SoftMax}_{\tau}(X\boldsymbol{a})\\
        &=X^T\operatorname{SoftMax}_{\tau}\bigl[(\hat{X}U^T+\boldsymbol{1}\bar{\boldsymbol{t}}^T)\boldsymbol{a}\bigr]\\
        &=X^T\operatorname{SoftMax}_{\tau}\bigl[\hat{X}U^TV\hat{\boldsymbol{a}}_{\text{full}}+\boldsymbol{1}\bar{\boldsymbol{t}}^T\boldsymbol{a}\bigr],
    \end{aligned}
\end{equation}
where $\boldsymbol{1}\in\mathbb{R}^{C}$ is a vector of all ones.  $\boldsymbol{a}=V\hat{\boldsymbol{a}}_{\text{full}}=V(\hat{\boldsymbol{a}}+\hat{\bar{\boldsymbol{a}}})$ is the decomposition of $\boldsymbol{a}$ including the mean. $\hat{\bar{\boldsymbol{a}}}$ is the projection of the audio mean component along the audio directions $\bar{\boldsymbol{a}}=V\hat{\bar{\boldsymbol{a}}}$. It can be seen as a prior to help predict the audio-text similarity even before the audio is known:
    \begin{align}
        \boldsymbol{t}\cdot\boldsymbol{a}&=(U\hat{\boldsymbol{t}}+U\hat{\bar{\boldsymbol{t}}})^T(V\hat{\boldsymbol{a}}+V\hat{\bar{\boldsymbol{a}}})\\
        &=\hat{\boldsymbol{t}}^TU^TV\hat{\boldsymbol{a}}+\hat{\boldsymbol{t}}^TU^TV\hat{\bar{\boldsymbol{a}}}+\hat{\bar{\boldsymbol{t}}}^TU^TV\hat{\boldsymbol{a}}+\hat{\bar{\boldsymbol{t}}}^TU^TV\hat{\bar{\boldsymbol{a}}}.\notag
    \end{align}
Note the fourth is a constant, and the second is this audio prior term.
In Eq.~(\ref{eq:actualpd_1}), $\boldsymbol{1}\bar{\boldsymbol{t}}^T\boldsymbol{a}$ is the same across dimensions and can be removed from softmax. 
In addition, the temperature $\tau$ is tiny. Thus, we approximate the softmax as a gating function that selects top similarity values and behaves linearly on the selected ones. Continuing from Eq.~(\ref{eq:actualpd_1}):
\begin{align}\label{eq:actualpd_2}
    \dots\!&= X^T\operatorname{SoftMax}_{\tau}(\hat{X}U^TV\hat{\boldsymbol{a}}_{\text{full}})\notag\\
    &\approx X^T_{\text{gated}}(k\hat{X}_{\text{gated}}U^TV\hat{\boldsymbol{a}}_{\text{full}}+C\boldsymbol{1})\notag\\
    &=(U\hat{X}^T_{\text{gated}}+\bar{\boldsymbol{t}}\boldsymbol{1}^T)(k\hat{X}_{\text{gated}}U^TV\hat{\boldsymbol{a}}_{\text{full}})+CX_{\text{gated}}^T\boldsymbol{1}\\
    &=kU\hat{X}^T_{\text{gated}}\hat{X}_{\text{gated}}U^TV\hat{\boldsymbol{a}}_{\text{full}}\!+\!k\bar{\boldsymbol{t}}(\boldsymbol{1}^T\hat{X}_{\text{gated}}U^TV\hat{\boldsymbol{a}}_{\text{full}})\!+\!C'\bar{\boldsymbol{t}}_{\text{g}}\notag\\
    &=kU\hat{X}^T_{\text{gated}}\hat{X}_{\text{gated}}U^TV\hat{\boldsymbol{a}}_{\text{full}}+k'\bar{\boldsymbol{t}}+C'\bar{\boldsymbol{t}}_{\text{g}}.\notag
\end{align}
\looseness=-1 We see a form similar to linear PD. Here, ${X}_{\text{gated}}\in\mathbb{R}^{N_{\text{gated}}\times C}$ is the selected text with high head similarity $\hat{X}U^TV\hat{\boldsymbol{a}}_{\text{full}}$. $\hat{X}_{\text{gated}}\in\mathbb{R}^{N_{\text{gated}}\times C}$ is the projection. 
$X^T_{\text{gated}}=U\hat{X}^T_{\text{gated}}+\bar{\boldsymbol{t}}\boldsymbol{1}^T$ is the decomposition. 
$k,C>0$ are softmax approximation coefficients. $X_{\text{gated}}^T\boldsymbol{1}=N_{\text{gated}}\bar{\boldsymbol{t}}_{\text{g}}$, where $\bar{\boldsymbol{t}}_{\text{g}}$ is the mean of the selected top-similarity embeddings $X_{\text{gated}}$. $k'=k(\bar{\tilde{\boldsymbol{t}}}_{\text{g}}^T\boldsymbol{a})$ and $C'=N_{\text{gated}}C$. $\bar{\tilde{\boldsymbol{t}}}_{\text{g}}$ is $\bar{\boldsymbol{t}}_{\text{g}}$ with text mean stripped. Now, the text mean $\bar{\boldsymbol{t}}$ and mean of selected embeddings $\bar{\boldsymbol{t}}_{\text{g}}$ are in $\boldsymbol{t}^{\text{pd}(\boldsymbol{a})}$. The rescaling now uses local covariance $\hat{X}^T_{\text{gated}}\hat{X}_{\text{gated}}$, computed on a small cluster of text neighbors tailored to each input. It rebuilds tails with local covariance instead of erasing them again after the gate $U^TV$.
\section{Decomposition and Retrieval for Other CLAP}\label{sec:app_decomp_ret_other_clap}
\begin{figure}
    \centering
    \subfloat[ASC-HTSAT-RoBERTa \cite{bai2025audiosetcaps} on Clotho.]{\includegraphics[width=0.9\linewidth]{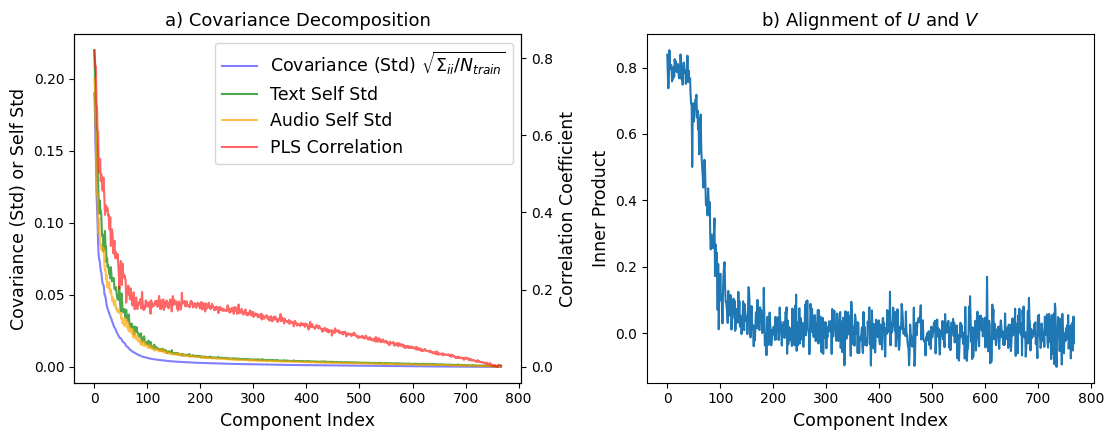}}\\
    \subfloat[ASC-HTSAT-RoBERTa \cite{bai2025audiosetcaps} on AudioCaps.]{\includegraphics[width=0.9\linewidth]{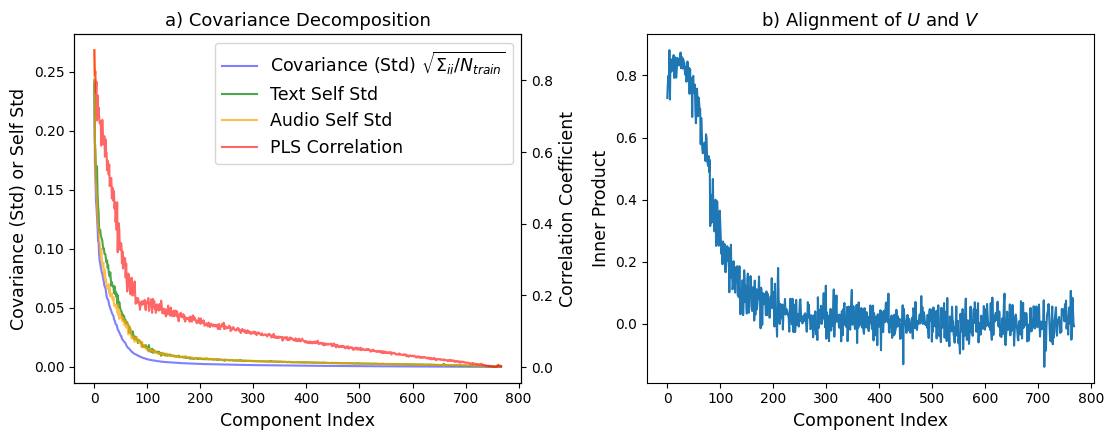}}\\
    \subfloat[CNN14-BERT-PT \cite{mei2024wavcaps} on Clotho.]{\includegraphics[width=0.9\linewidth]{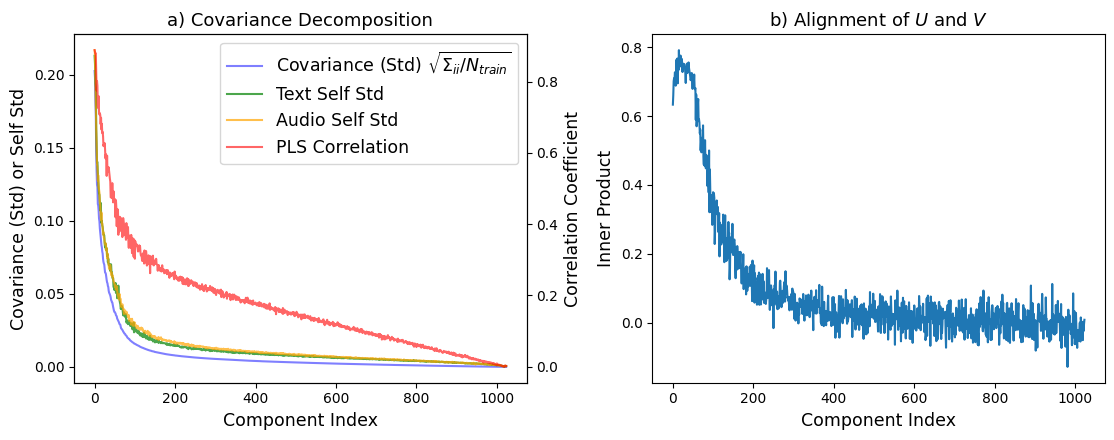}}\\
    \subfloat[CNN14-BERT-PT \cite{mei2024wavcaps} on AudioCaps.]{\includegraphics[width=0.9\linewidth]{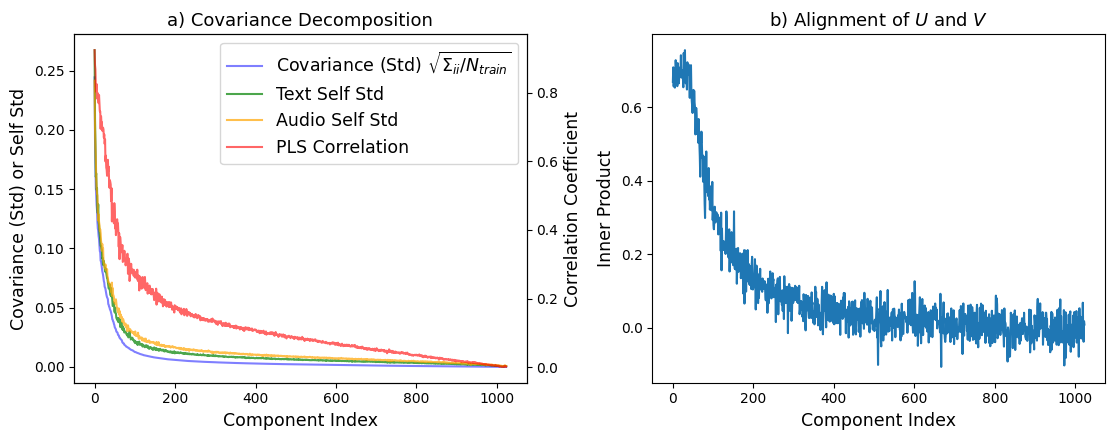}}\\
    \subfloat[CLAP used by DRCap \cite{li2025drcap} on Clotho.]{\includegraphics[width=0.9\linewidth]{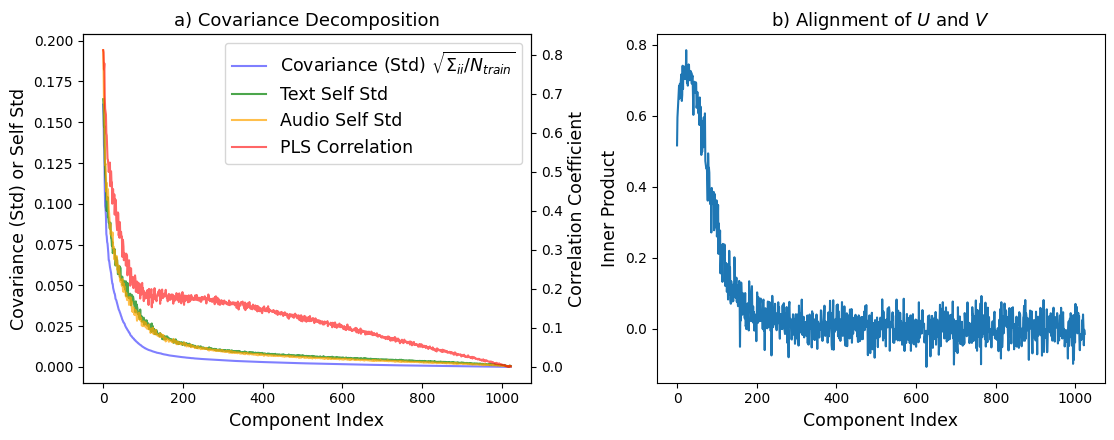}}\\
    \subfloat[CLAP used by DRCap \cite{li2025drcap} on AudioCaps.]{\includegraphics[width=0.9\linewidth]{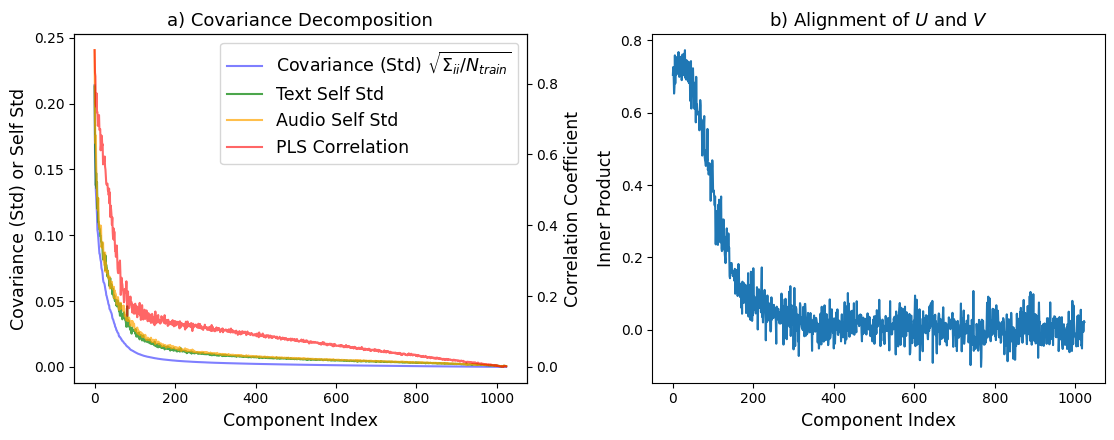}}\\
    \caption{PLS covariance decomposition and UV alignment for other CLAP.}
    \label{fig:s_decomp_uvotherclap}
\end{figure}
We show decomposition (in Fig.~\ref{fig:s_decomp_uvotherclap}) and text-to-audio retrieval (in Table~\ref{tab:t2aotherclap}) results with other CLAP models, which are ASC-HTSAT-RoBERTa \footnote{ATR\_HTSAT+RoBERTa\_ASC\_PT\_mask25\_T2AR1\_32.8\_A2TR1\_44.5.pt} pretrained on Audio\-Set\-Caps \cite{bai2025audiosetcaps}, CNN14-BERT-PT pretrained on WavCaps+AudioCaps+Clotho \cite{mei2024wavcaps}, and the CLAP model used in DRCap \cite{li2025drcap} pretrained on WavCaps+SoundVECaps\cite{yuan2025sound}. 
In Table~\ref{tab:t2aotherclap}, ``Cl'' means Clotho, ``AC'' means AudioCaps, ``ASC'' means CLAP pretrained on AudioSetCaps, ``mAP'' means mAP10, ``PLSHd'' (or ``PLSHdW'') means our method PLSHead (or PLSHeadW), using top 100 PLS-SVD concept directions. 
\begin{table}[tbp]
    \centering
    \caption{Text-to-Audio Retrieval with Top 100 Projections.}
    \setlength\tabcolsep{1.6pt}
    \label{tab:t2aotherclap}
    \begin{tabular}{lccccccc}\toprule
         CLAP (Dataset) Method & R1 & R5 & R10 & R50 & MeanR & MedR & mAP \\\midrule
         ASC (Cl) Raw & 10.89 & 29.51 & 40.82 & 73.09 & 50.02 & \underline{17.00} & 18.90 \\
         ASC (Cl) PLSHd  & \textbf{13.11} & \underline{33.32} & \textbf{46.43} & \textbf{77.53} & \textbf{43.53} & \textbf{13.00} & \textbf{22.05} \\
         ASC (Cl) PLSHdW & \underline{13.03} & \textbf{33.34} & \underline{46.39} & \underline{77.49} & \underline{43.58} & \textbf{13.00} & \underline{22.02} \\\midrule
         ASC (AC) Raw & 32.83 & \underline{66.54} & 79.52 & 95.86 & 10.07 & \textbf{3.00} & 47.06 \\
         ASC (AC) PLSHd & \underline{33.61} & \textbf{68.13} & \underline{80.67} & \underline{96.20} & \underline{9.92} & \textbf{3.00} & \underline{48.30} \\
         ASC (AC) PLSHdW & \textbf{34.15} & \textbf{68.13} & \textbf{80.92} & \textbf{96.24} & \textbf{9.75} & \textbf{3.00} & \textbf{48.67} \\\midrule
         WCCNN14 (Cl) Raw & 21.11 & 46.22 & \underline{59.46} & 85.68 & 29.69 & \textbf{7.00} & 31.94 \\
         WCCNN14 (Cl) PLSHd & \textbf{21.72} & \underline{46.32} & \underline{59.46} & \textbf{86.85} & \textbf{27.27} & \textbf{7.00} & \textbf{32.25} \\
         WCCNN14 (Cl) PLSHdW & \underline{21.49} & \textbf{46.58} & \textbf{59.50} & \underline{86.83} & \underline{27.31} & \textbf{7.00} & \underline{32.23} \\\midrule
         WCCNN14 (AC) Raw & \textbf{34.44} & \textbf{69.09} & \textbf{82.49} & \textbf{97.24} & \textbf{8.59} & \textbf{3.00} & \textbf{49.09} \\
         WCCNN14 (AC) PLSHd & 32.81 & 67.69 & 79.54 & 96.68 & 9.65 & \textbf{3.00} & 47.50 \\
         WCCNN14 (AC) PLSHdW & \underline{33.42} & \underline{68.07} & \underline{79.85} & \underline{96.76} & \underline{9.56} & \textbf{3.00} & \underline{47.95} \\\midrule
         DRCap (Cl) Raw & 17.88 & 41.70 & 55.04 & 82.72 & 37.78 & \textbf{8.00} & 28.17 \\
         DRCap (Cl) PLSHd & \textbf{18.33} & \textbf{42.37} & \textbf{55.87} & \textbf{84.63} & \textbf{33.69} & \textbf{8.00} & \textbf{28.56} \\
         DRCap (Cl) PLSHdW & \underline{18.18} & \underline{42.30} & \underline{55.52} & \underline{84.59} & \underline{33.79} & \textbf{8.00} & \underline{28.41} \\\midrule
         DRCap (AC) Raw & 33.29 & 65.94 & 79.50 & 95.80 & 10.61 & \textbf{3.00} & 47.40 \\
         DRCap (AC) PLSHd & \underline{33.40} & \underline{67.44} & \underline{80.44} & \underline{96.43} & \underline{9.60} & \textbf{3.00} & \underline{47.74} \\
         DRCap (AC) PLSHdW & \textbf{33.83} & \textbf{67.80} & \textbf{80.90} & \textbf{96.57} & \textbf{9.48} & \textbf{3.00} & \textbf{48.15}
         \\\bottomrule
    \end{tabular}
\end{table}

\ifCLASSOPTIONcaptionsoff
  \newpage
\fi

\bibliographystyle{IEEEtran}
\bibliography{IEEEabrv,paper.bib}

\end{document}